\documentclass[]{pasj02} 
\usepackage[switch,mathlines]{lineno} 

\usepackage{xcolor}

\jyear{2024}
\Received{}
\Accepted{}

\begin{document} 

\title{Measuring the asymmetric expansion of the Fe ejecta of Cassiopeia A with XRISM/Resolve}

\author{
Aya \textsc{Bamba},\altaffilmark{1,2,3}\orcid{0000-0003-0890-4920}
\email{bamba@phys.s.u-tokyo.ac.jp}
Manan \textsc{Agarwal},\altaffilmark{4}\orcid{0000-0001-6965-8642}
Jacco \textsc{Vink},\altaffilmark{4,5}\orcid{0000-0002-4708-4219}
Paul \textsc{Plucinsky},\altaffilmark{6}\orcid{0000-0003-1415-5823}
Yukikatsu \textsc{Terada},\altaffilmark{7,8}\orcid{0000-0002-2359-1857}
Ehud \textsc{Behar},\altaffilmark{9}\orcid{0000-0001-9735-4873}
Satoru \textsc{Katsuda},\altaffilmark{7}\orcid{0000-0002-1104-7205} 
Koji \textsc{Mori},\altaffilmark{10}\orcid{0000-0002-0018-0369}
Makoto \textsc{Sawada},\altaffilmark{11}\orcid{0000-0003-2008-6887}
Hironori \textsc{Matsumoto},\altaffilmark{12} 
Lia \textsc{Corrales},\altaffilmark{13}\orcid{0000-0002-5466-3817} 
Adam \textsc{Foster},\altaffilmark{14}\orcid{0000-0003-3462-8886}
Shin-ichiro \textsc{Fujimoto},\altaffilmark{15}\orcid{0000-0002-7273-2740}
Liyi \textsc{Gu},\altaffilmark{5}\orcid{0000-0001-9911-7038}
Kazuhiro \textsc{Ichikawa},\altaffilmark{10}
Kai \textsc{Matsunaga},\altaffilmark{16}\orcid{0009-0003-0653-2913} 
Tsunefumi \textsc{Mizuno},\altaffilmark{17}\orcid{0000-0001-7263-0296}
Hiroshi \textsc{Murakami},\altaffilmark{18}\orcid{0000-0002-3844-5326}
Hiroshi \textsc{Nakajima},\altaffilmark{19}\orcid{0000-0001-6988-3938}
Toshiki \textsc{Sato},\altaffilmark{20}\orcid{0000-0001-9267-1693}
Haruto \textsc{Sonoda},\altaffilmark{8} 
Shunsuke \textsc{Suzuki},\altaffilmark{8,21} 
Dai \textsc{Tateishi},\altaffilmark{1}\orcid{0000-0003-0248-4064}
Hiroyuki \textsc{Uchida},\altaffilmark{16}\orcid{0000-0003-1518-2188}
Masahiro \textsc{Ichihashi},\altaffilmark{1}\orcid{0000-0001-7713-5016}
Kumiko \textsc{Nobukawa},\altaffilmark{22}\orcid{0000-0002-0726-7862}
Salvatore \textsc{Orlando}\altaffilmark{23}\orcid{0000-0003-2836-540X}
}
\altaffiltext{1}{Department of Physics, Graduate School of Science,
The University of Tokyo, 7-3-1 Hongo, Bunkyo-ku, Tokyo 113-0033, Japan}
\altaffiltext{2}{Research Center for the Early Universe, School of Science, The University of Tokyo, 7-3-1
Hongo, Bunkyo-ku, Tokyo 113-0033, Japan}
\altaffiltext{3}{Trans-Scale Quantum Science Institute, The University of Tokyo, Tokyo  113-0033, Japan}
\altaffiltext{4}{Anton Pannekoek Institute/GRAPPA, University of Amsterdam, Science Park 904, 1098 XH Amsterdam, The Netherlands}
\altaffiltext{5}{SRON Netherlands Institute for Space Research, Niels Bohrweg 4, 2333 CA Leiden, The Netherlands}
\altaffiltext{6}{Harvard-Smithsonian Center for Astrophysics, MS-3, 60 Garden Street, Cambridge, MA, 02138, USA}
\altaffiltext{7}{Graduate School of Science and Engineering, Saitama University, 255 Shimo-Ohkubo, Sakura, Saitama 338-8570, Japan}
\altaffiltext{8}{ISAS/JAXA, 3-1-1 Yoshinodai, Chuo-ku, Sagamihara, Kanagawa 252-5210, Japan}
\altaffiltext{9}{Department of Physics, Technion, Technion City, Haifa 3200003, Israel}
\altaffiltext{10}{Faculty of Engineering, University of Miyazaki, Miyazaki 889-2192, Japan}
\altaffiltext{11}{Department of Physics, Rikkyo University, Tokyo 171-8501, Japan}
\altaffiltext{12}{Department of Earth and Space Science, Osaka University, Osaka 560-0043, Japan}
\altaffiltext{13}{Department of Astronomy, University of Michigan, MI 48109, USA}
\altaffiltext{14}{Center for Astrophysics, Harvard-Smithsonian, MA 02138, USA}
\altaffiltext{15}{National Institute of Technology Kumamoto College, 2659-2 Suya, Koshi, Kumamoto 861-1102, Japan}
\altaffiltext{16}{Department of Physics, Kyoto University, Kyoto 606-8502, Japan}
\altaffiltext{17}{Department of Physics, Hiroshima University, Hiroshima 739-8526, Japan}
\altaffiltext{18}{Department of Data Science, Tohoku Gakuin University, Miyagi 984-8588, Japan}
\altaffiltext{19}{College of Science and Engineering, Kanto Gakuin University, Kanagawa 236-8501, Japan}
\altaffiltext{20}{School of Science and Technology, Meiji University, Kanagawa, 214-8571, Japan}
\altaffiltext{21}{Department of Physical Sciences, Aoyama Gakuin University, 5-10-1 Fuchinobe, Sagamihara, Kanagawa 252-5258, Japan; Institute of Laser Engineering, Osaka University, 2-6, Yamadaoka, Suita, Osaka 565-0871, Japan}
\altaffiltext{22}{Department of Science, Faculty of Science and Engineering, KINDAI University, Osaka 577-8502, Japan}
\altaffiltext{23}{INAF—Osservatorio Astronomico di Palermo, Piazza del Parlamento 1, I-90134 Palermo, Italy}



\maketitle

\begin{abstract}
The expansion structure of supernova remnants (SNRs) is important for understanding not only how heavy elements are distributed into space, but also how supernovae explode. The ejecta expansion structure of the young core-collapse SNR Cas A is investigated, with Doppler parameter mapping of the Fe-K complex by the Resolve microcalorimeter onboard the X-ray Imaging and Spectroscopy Mission, XRISM. 
It is found that the Fe ejecta are blueshifted in the southeast (SE) and redshifted in the northwest (NW), indicating an incomplete shell structure, similar to the intermediate mass elements (IMEs), such as Si and S. 
The Fe has a velocity shift of $\sim1400$ km~s$^{-1}$ in the NW and $\sim2160$ km~s$^{-1}$ in the SE region, with the error range of a few 100s km~s$^{-1}$.
These values are consistent with those for the IMEs in the NW region, whereas larger than those for the IMEs in the SE region,
although the large error region prevented us from concluding which component has significantly higher velocity.
The line broadening is larger in the center with values of $\sim$2000--3000~km~s$^{-1}$, and smaller near the edges of the remnant.
The radial profiles of the Doppler shift and broadening of the IMEs and Fe indicate that the Fe ejecta may expand asymmetrically as IME ejacta,
although the large error regions do not allow us to conclude it.
Moreover, we see little bulk Doppler broadening of the Fe lines in the northeastern jet region whereas the IME lines exhibit significant broadening.
No such narrow lines are detected in the NW region.
These findings suggest an asymmetric expansion of the ejecta potentially driven by large-scale asymmetries originating from the supernova explosion. This interpretation aligns with the large-scale asymmetries predicted by models of neutrino-driven supernova explosions.
\end{abstract}


\section{Introduction}

Supernovae and their remnants (supernova remnants; SNRs) play crucial roles in the chemical evolution of the Universe by providing heavy elements to interstellar medium.
Measuring the expansion of the explosion ejecta will not only reveal how these heavy elements are ejected into interstellar space, but also provide insight into the explosion mechanism of the supernova explosion itself,
since it is now believed that multi-dimensional
effects are essentially important in the core-collapse supernova explosion mechanism
\citep[for review]{maeda2022}.

Cassiopeia A (Cas A) is one of the best objects to study young core-collapse SNRs.
The remnant has a radius of $\sim$2.8~arcmin,
which corresponds to 2.8~pc at the distance of 3.4~kpc \citep{reed1995}.
The thermal X-ray
emission from Cas A is dominated by the shocked ejecta.
Many attempts have been made to study the expansion of this SNR.
The forward- and reverse-shock expansion measurements have been done
with proper motion \citep{vink2022}
in the synchrotron X-rays from the shock fronts
\citep{vink2003,bamba2005}
and optical flow method in the synchrotron-dominant energy band and the Si-K line band \citep{sato2018}.
\citet{willingale2002} showed that
intermediate mass elements (IMEs, such as Si, S, and Ar), and Fe have different spatial distributions as well as Doppler shifts from each other
(see also \cite{lazendic2006,delaney2010,sato2021}). 
\citet{orlando2022} performed three-dimensional (magneto)-hydrodynamic simulations of evolution from the supernova to its interaction with a massive circumstellar shell,
and succeeded to reproduce asymmetric structure of the reverse shock of Cas A, and derived the progenitor information.
For smaller scales, \citet{wongwathanarat2015} also made three-dimensional hydrodynamic simulations of the evolution of core-collapse supernovae, and found that heavy elements can be ejected by Rayleigh-Taylor instabilities at the C+O/He and He/H composition-shell interfaces after the passage of the SN shock.
\citet{sakai2024} has made multiepoch maximum likelihood estimation approach to Chandra images of Cas~A to measure the proper motion of emission components, and found that one of the component interacts with circumstellar material.

On the other hand, the three dimensional structure of the hot plasma has not been well studied using Doppler measurements. This is especially true for Fe.
It is due to the limited spectral resolution of previous X-ray missions
for extended objects.
\citet{willingale2002} has made the Doppler-shift map of Fe with XMM-Newton data and showed that some Fe knots show a very large Doppler shift.
This should be tested by a new spectrometer with excellent energy resolution, $\sim$4~eV,
since the Fe in the ejecta is still ionizing and 
the profile of the Fe-K complex should be complicated due to contributions from ions of low-ionization states.
Note that, in contrast, the three-dimensional structure has already been studied for cold plasma observed with HST (e.g., \cite{delaney2010}; \cite{milisavljevic2013}).
Doing this for the hot plasma is a necessary complement for connecting features observed in the main shell of shocked ejecta with the features observed by HST (and, more recently, by JWST).
This will allow to reconstruct the whole structure of the ejecta.

In this paper, we for the first time derive Doppler parameters of Fe from the ejecta of Cas A
with the excellent energy resolution of Resolve \citep{ishisaki2022}
onboard the X-ray Imaging and Spectroscopy Mission (XRISM; \citep{tashiro2022}).
We make comparisons with the Doppler parameters of IME taken with Resolve, for which the results  are reported by \citet{vink2024} in this volume.
In Section~\ref{sec:obs}, we detail observation methods and data reduction. Section~\ref{sec:analysis} outlines our imaging and spectral analysis results, while Section~\ref{sec:discuss} offers a discussion of our findings.

\section{Observations and data reduction}
\label{sec:obs}

Cas A has been observed with XRISM
with two pointings, southeast (ObsID: 000129000, ``SE" hereafter) and northwest (ObsID: 000130000, ``NW" hereafter).
Since we concentrate on studying Doppler parameters of Fe-K lines, we use the Resolve data only.
We reprocessed the data with the \texttt{xapipeline} task \citep{terada2021,lowenstein2020} using the calibration database version 20240815 with the standard screening criteria.\footnote{ https://heasarc.gsfc.nasa.gov/docs/xrism/analysis/abc\_guide/XRISM\_Data\_Analysis.html}
In addition to that, we applied the Resolve rise-time screening in the XRISM quick start guide.\footnote{https://heasarc.gsfc.nasa.gov/docs/xrism/analysis/quickstart/index.html}
It is known that the aim point shows a small periodic excursion from the nominal position depending on the base-panel temperature of the spacecraft \citep{kanemaru2024,kanemaru2025}.
We checked the impact of this effect by comparing results with and without an additional cut on the pointing stability and found no significant change. Therefore, we present results without applying the pointing stability cut.
The resultant exposure times are 181.3~ks for SE and 166.6~ks for NW.
We selected only high-primary events.
The data analysis was done with HEASoft version 6.34.
Throughout this paper, 
we quote errors at the 90\% confidence level.

\section{Analysis and results}
\label{sec:analysis}

In this study, we selected two emission features,
lines from He-like Si and S as the tracer of IMEs and K-shell lines from Fe as the tracer of Fe.
The IME Doppler parameters are studied by \citet{vink2024} in this volume, which we adopt for comparison to the Fe results studied in this paper.

\subsection{Line profiles}

\begin{figure*}
 \begin{center}
  \includegraphics[width=8cm]{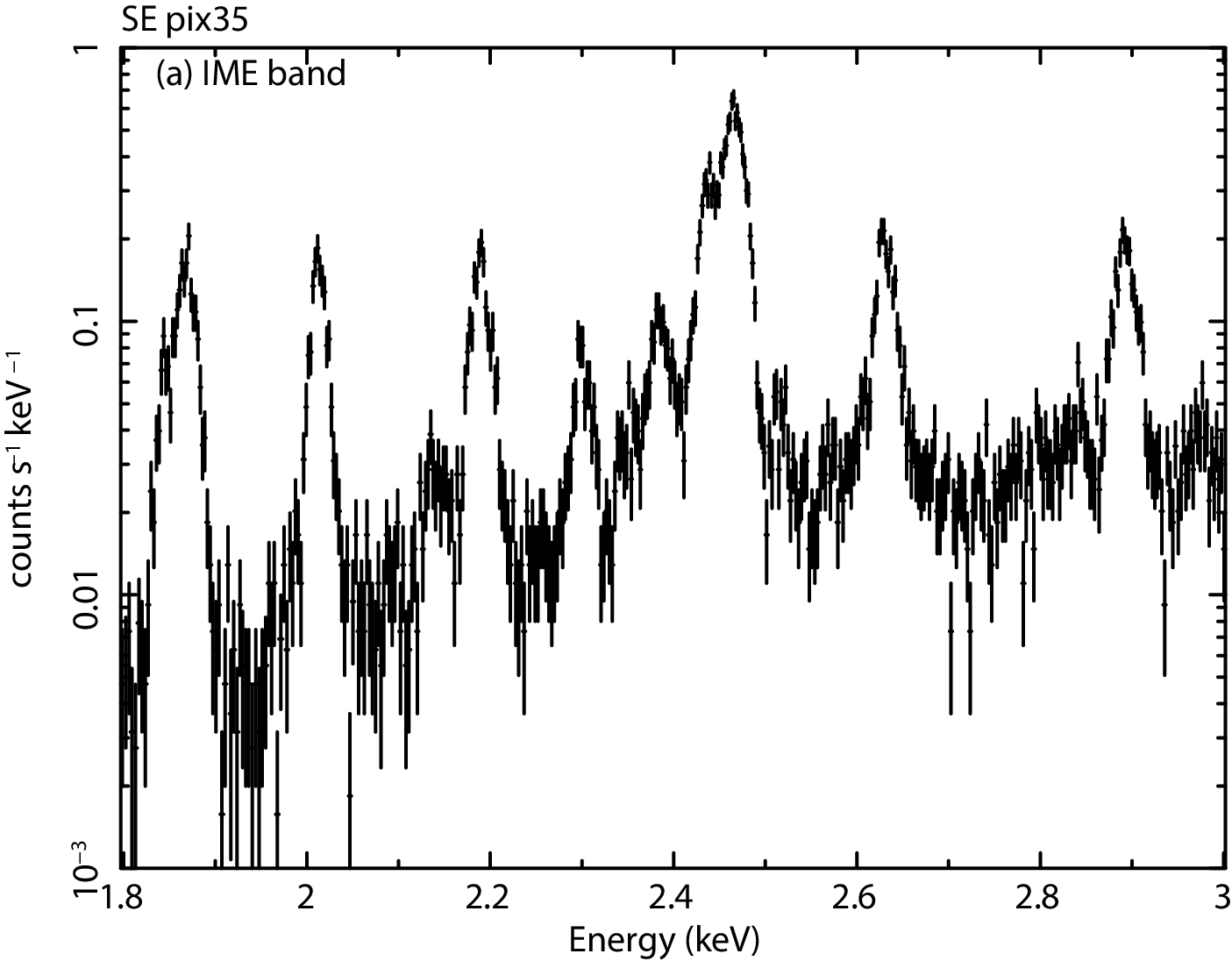} 
  \includegraphics[width=8cm]{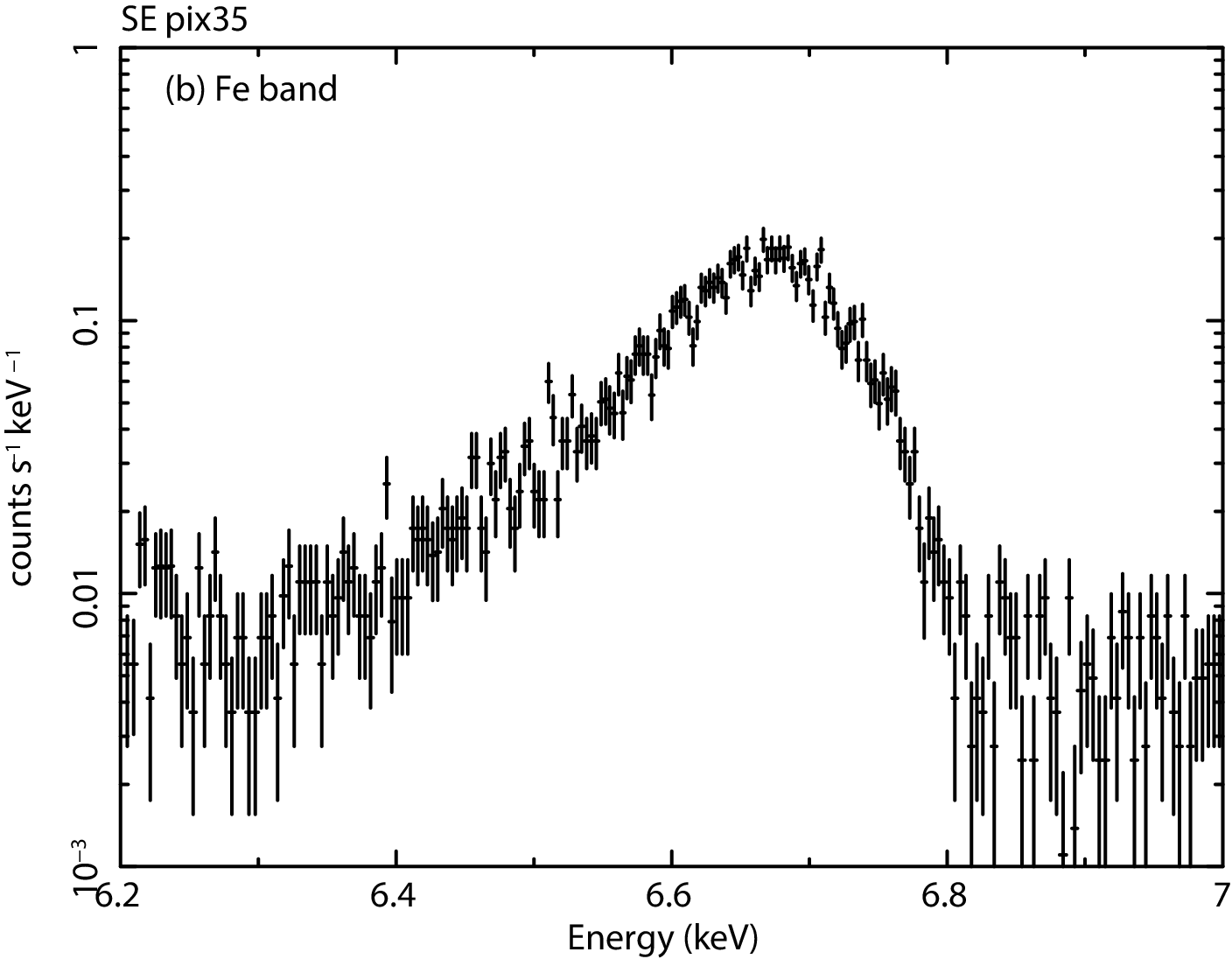} 
\end{center}
\caption{IME (a) and Fe-K (b) band spectra of pixel 35 near the center of the Resolve detector array in the SE observation.  
 {Alt text: Two line graphs showing IME and Fe-K line band spectra. The holizontal axis in each graph show the energy in keV, whereas the vertical one is plotted with the unit of counts per sec per keV.} 
}\label{fig:lineexample}
\end{figure*}

Figure~\ref{fig:lineexample} shows a pair of examples of a single-pixel spectrum in IME (a) and the Fe-K (b) bands
(pixel 35 in the SE observation; see \cite{plucinsky2024}), see also Figure~\ref{fig:v-w}.
One can see broadened lines in the both bands, but the detailed line structure is different from each other.
The IME lines show basically symmetric structure, whereas the Fe line is asymmetrical with a tail extending to lower energies.
The likely cause of this asymmetry is that Fe is
further out of ionization equilibrium than the IMEs and
has a broad range of ionization states, resulting in many Fe lines from charge states lower than the He-like state, contributing to 
the Fe-K complex.
The range of charge states is too broad to be modeled with a single
non-equilibrium ionization model (such as {\tt nei} in Xspec).
In particular a single {\tt nei} model fails to reproduce the low energy tail of Fe with low ionization states. Instead the plasma must be characterized by a broad
range of ionization ages, which we found is well approximated by the
plane-parallel shock model ({\tt pshock} in Xspec) \citep{borkowski2001}.

\subsection{Fittings}

The spectral fitting to derive the Doppler parameters of Fe was performed based on
the C-statistic \citep{cash1979,kaastra2017},
using the Xspec software 
version 12.14.1 \citep{arnaud1996}
with the AtomDB version 3.0.9 \citep{foster2012}.
The spectra were binned with the optimal binning method \citep{kaastra2016} by the task {\tt ftgrouppha}.
We used 
6.2--7~keV band spectra for the fitting.
The response matrix of each pixel was created using {\tt rslmkrmf} task.
We used the auxiliary response file for an on-axis point source provided by the XRISM team,
since we fit the spectra in a narrow band and we do not care about the absolute flux.
For the fitting model, {\tt vpshock} model was applied
with the free parameters of electron temperature,
ionization time scale, Fe abundance, normalization,
and the redshift in the {\tt vpshock} model.
We represented Doppler broadening with a multiplicative model {\tt gsmooth}.
This model performs a spectral broadening
by redistributing photon flux in each energy bin defined by the energy grid in the redistribution matrix file
(0.5~eV step for Resolve).
Therefore, it is not optimal for applying to sharp spectral features comparable to the energy step size as it could produce artifacts in the broadened spectrum.
However, our choice is fine because the broadening in the Fe-K band is 40--60~eV as in Figure~1 (b),
much larger than the energy step size.
There can be a significant contamination of synchrotron X-rays especially in the Fe-K band
\citep{helder2008}.
We thus made another set of the fitting in the Fe-K band,
with a smoothed {\tt vpshock} model plus a power-law component
representing the X-ray synchrotron radiation component.
The photon index of the power-law component was fixed to be 3.2 \citep{helder2008}.
The only additional free parameter was the normalization of the power-law component.
The best-fit parameters for the Doppler effect with or without the power-law component do not change significantly within the range of their errors; thus, we just adopted the model with the power-law component.

The fitting returned reasonable C-statistic values for the all pixels without large systematic residuals,
and reasonable temperatures and ionization time scales ($n_et$) were obtained although the error ranges are large.
The best-fit redshift and line broadening scatter from $-9\times 10^{-3}$ to $+1.3\times 10^{-2}$ and 7.5~eV to 67~eV, respectively.
This is larger than
the calibration uncertainty of the gain of Resolve ($\sim\pm$1~eV below 9~keV; \cite{eckart2024,porter2024})
and that of the energy resolution 
(less than 0.3~eV below 10~keV\footnote{Available at https://heasarc.gsfc.nasa.gov/docs/xrism/calib/xrism\_caldbdocs.html}),
implying that XRISM/Resolve successfully showed that
Cas A shows spatially dependent redshift and Doppler broadening.
Note that the recent update on AtomDB (from version 3.0.9 to version 3.1.0) does not affect our results significantly.
With remaining thermal parameters,
we made the consistency check with previous Chandra results \citep{hwang2012}.
We found that the derived $n_et$ map is consistent with the map derived by Chandra \citep{hwang2012}.
This shows that we sucessfully derived spatially resolved Doppler parameters with thermal models even with narrow energy band fitting.
More detailed description on this comparison is summarized in section \ref{sec:nt}.

\subsection{Doppler maps}

\begin{figure*}
 \begin{center}
\includegraphics[width=8cm]{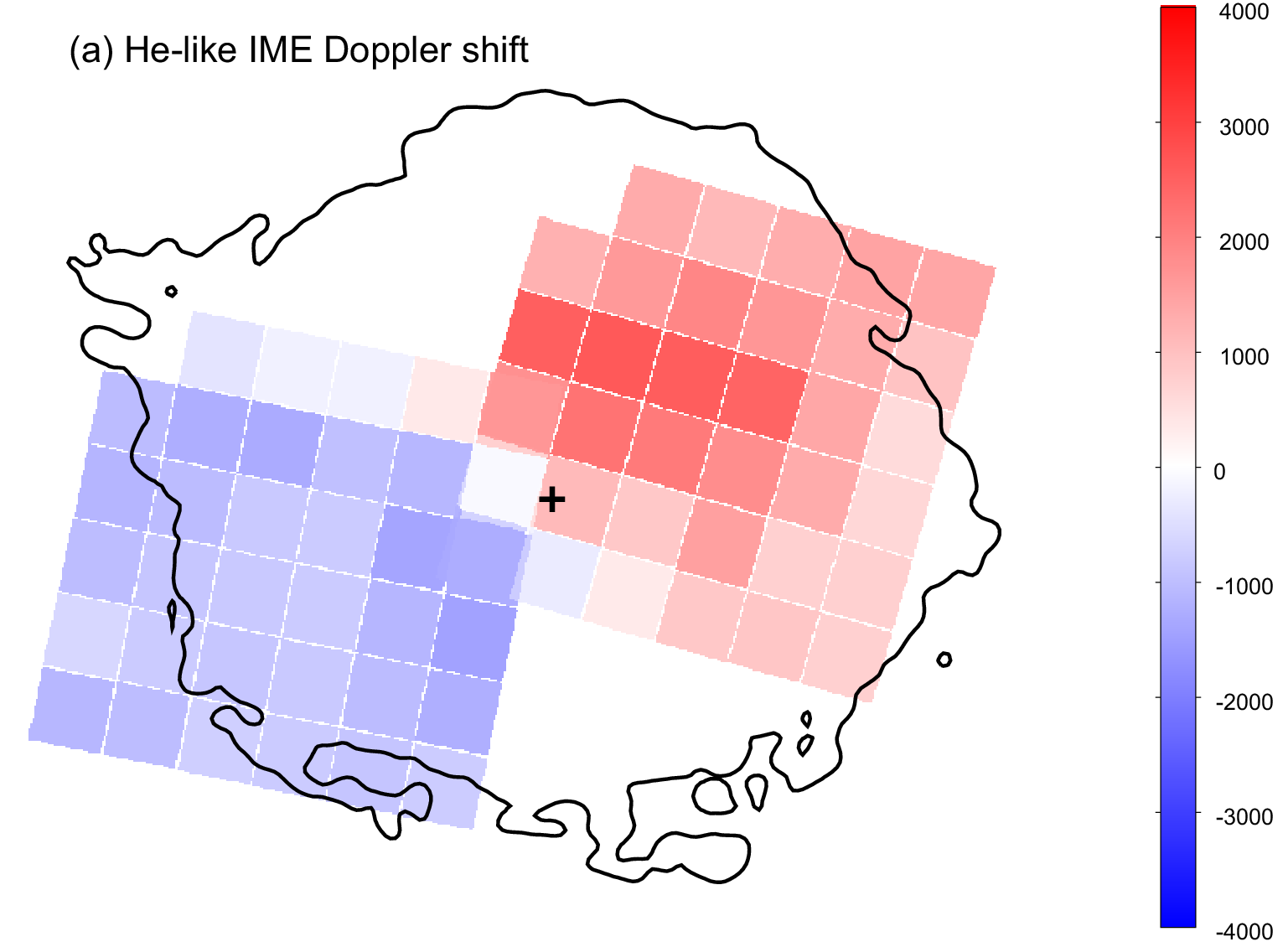}
\includegraphics[width=8cm]{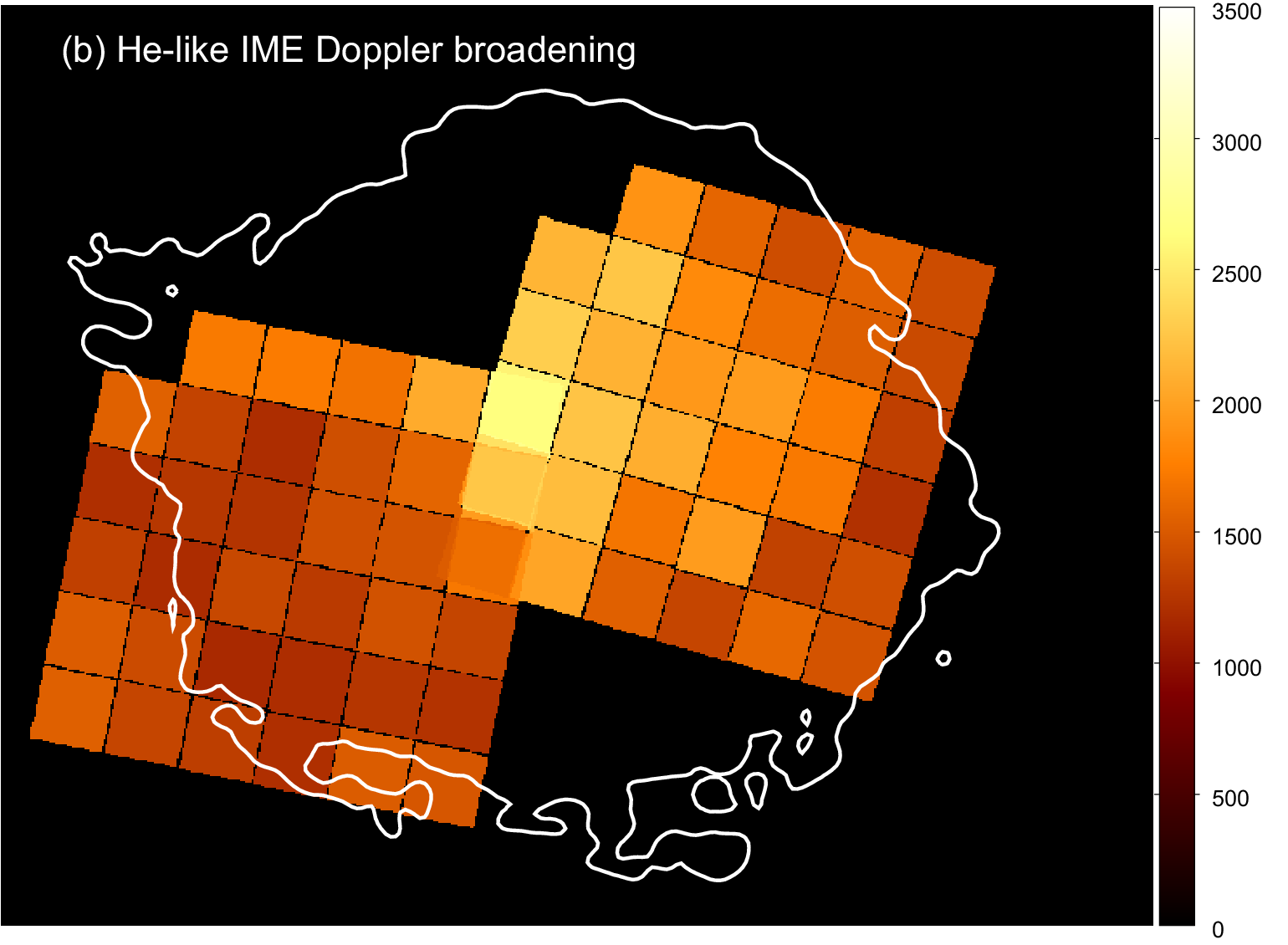}
\includegraphics[width=8cm]{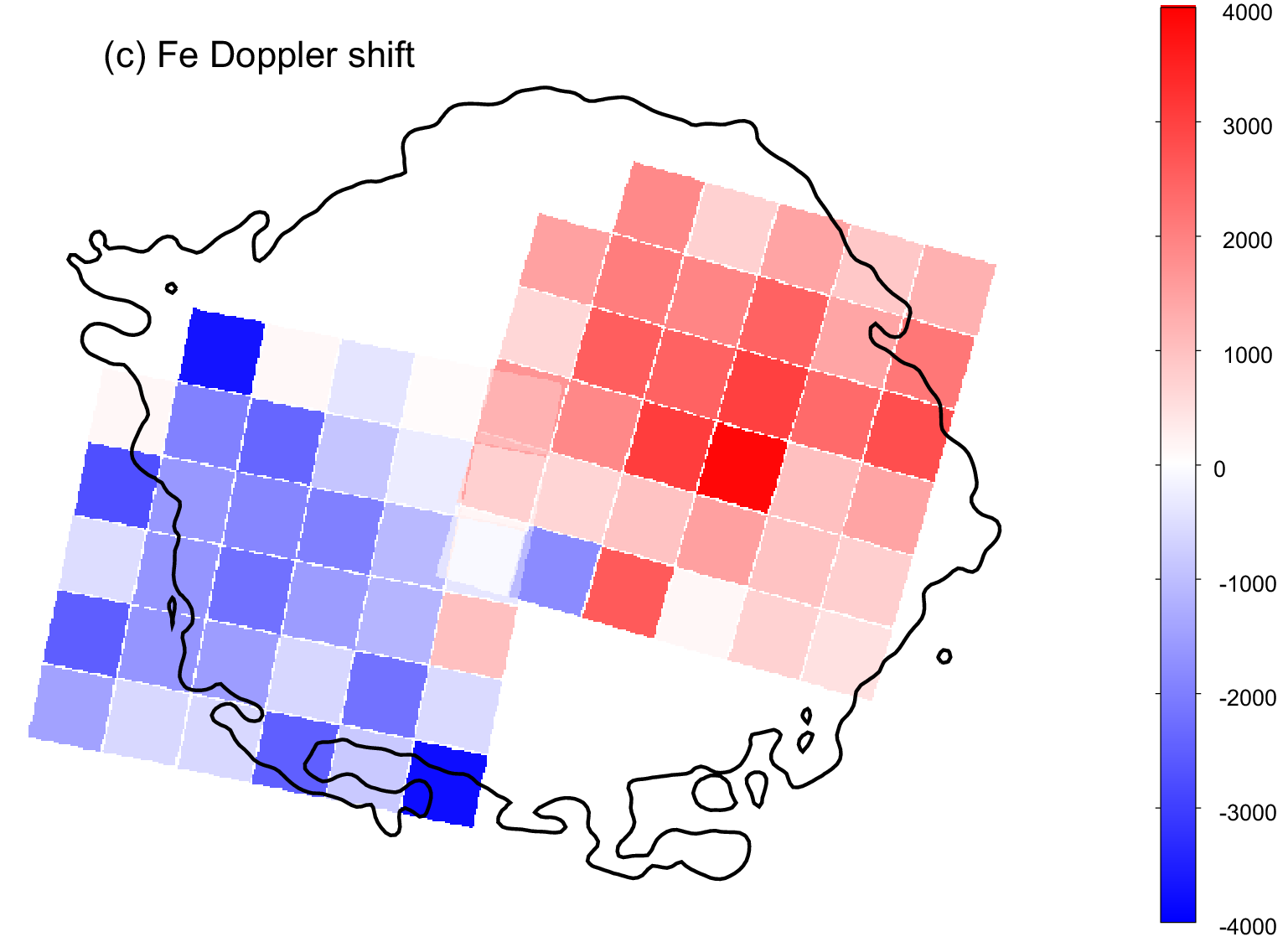}
\includegraphics[width=8cm]{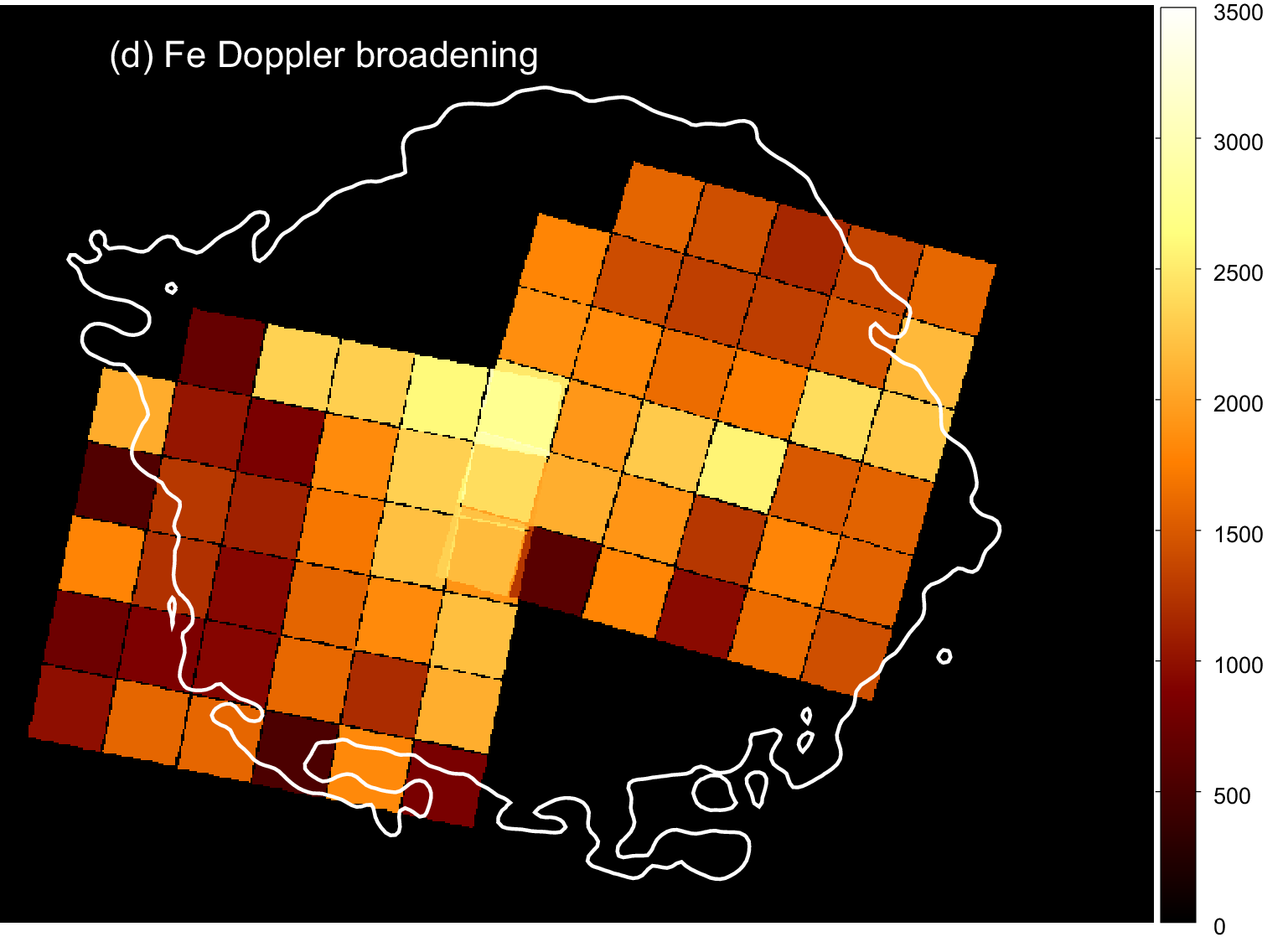}
\includegraphics[width=8cm]{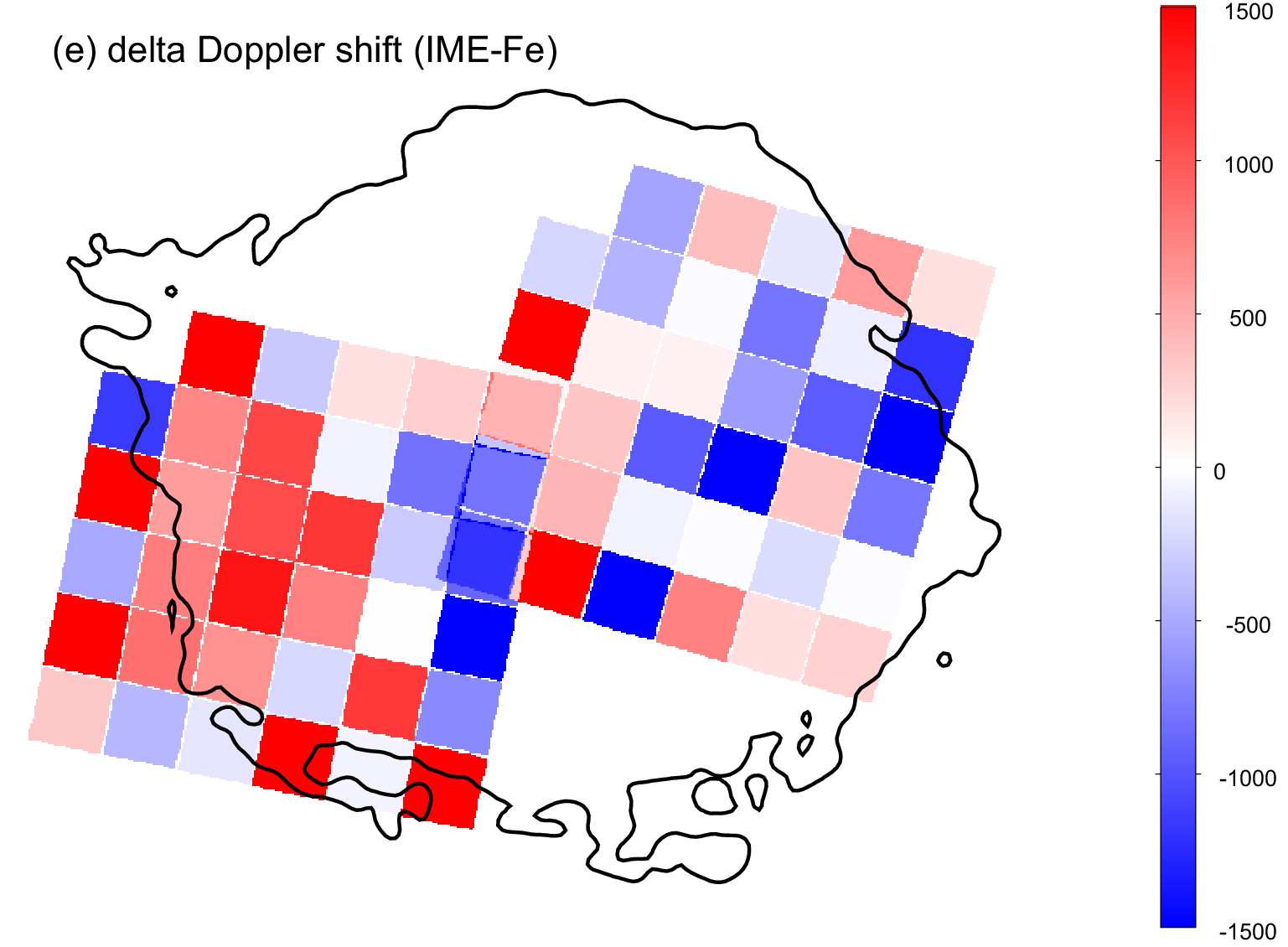}
\includegraphics[width=8cm]{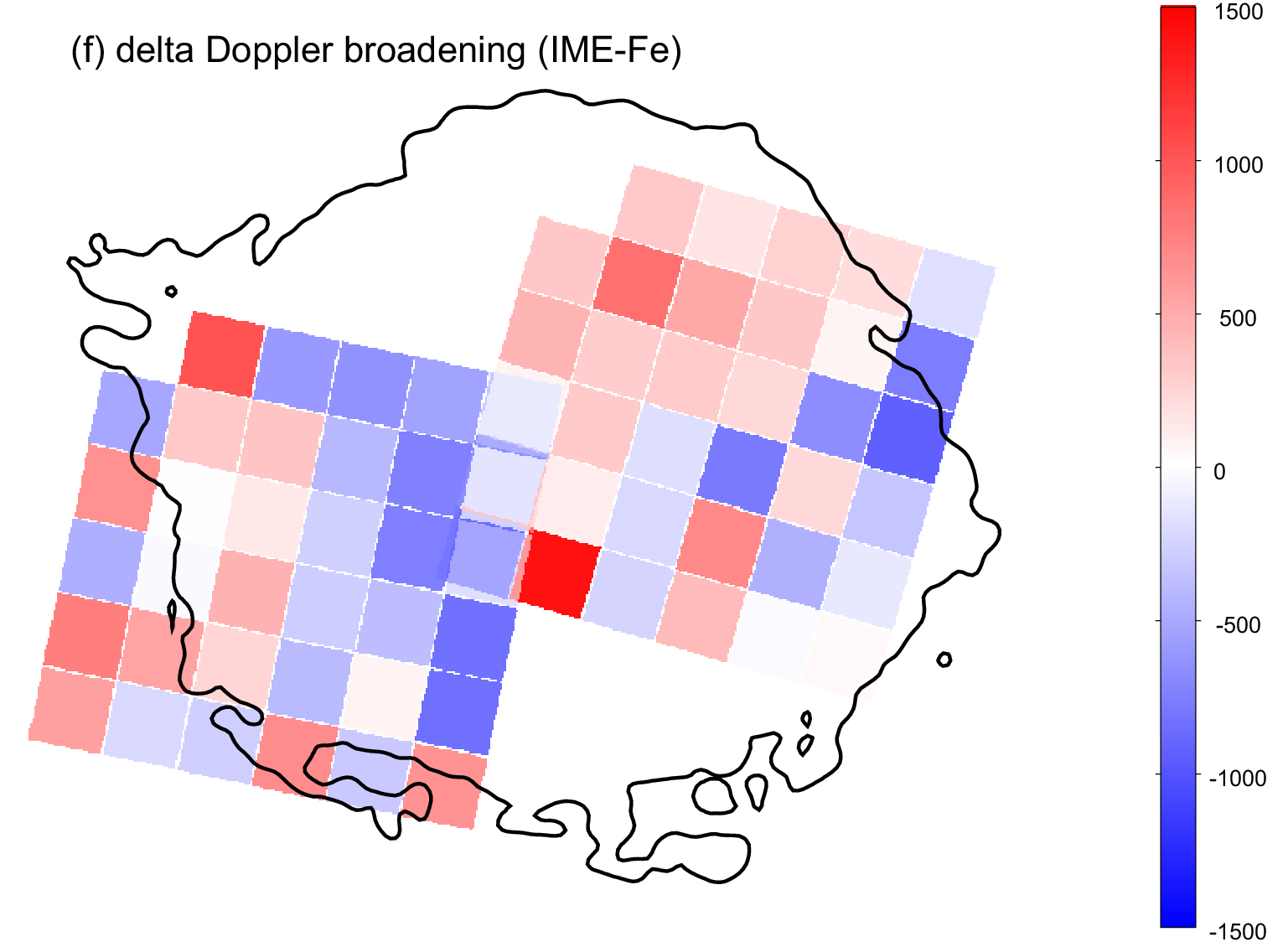}
\end{center}
\caption{Maps of (a) Doppler shift of He-like IME,
(b) Doppler broadening of He-like IME,
(c) Doppler shift of Fe-K,
(d) Doppler broadening of Fe-K.  
(e) Doppler shift difference of IME and Fe,
and (f) Doppler broadening difference of IME and Fe.
Contours show the Chandra whole-band image.
The unit of color scale is in km~s$^{-1}$ for the all panels.
Black + in the panel (a) represents the expansion center from \citet{thorstensen2001}.
{Alt text: Six maps of Doppler parameters of He-like IME and Fe-K lines.} 
}\label{fig:maps}
\end{figure*}

Figure~\ref{fig:maps} shows the Doppler shift and Doppler broadening maps
of He-like IME and Fe-K.
Note that we adopt the results of the He-like IME by \citet{vink2024},
which was derived from Resolve spectra of He-like Si and S.
Here, we converted the redshift and broadening to the velocity with the assumption that the Fe-K line has a centroid of 6.6~keV.
Note that the statistical error is large and the values are scattered
on the edge of the remnant.

Figure~\ref{fig:maps} (a) and (c) show the Doppler shift of He-like IME and Fe, respectively.
Both maps are quite similar, showing blue-shift in the SE region and red-shift in the NW region.
This implies that the explosion was not isotropic.
The maximum velocity shift is $\sim$4000~km~s$^{-1}$ for the both elements.

Figure~\ref{fig:maps} (b) and (d) are 
Doppler broadening maps of IME and Fe, respectively.
One can see that both maps are similar again,
showing that
the central part has a wider velocity dispersion than the peripheral part,
implying spherical expansion structure.
The maximum Doppler velocity broadening is $\sim$3000~km~s$^{-1}$
for the both elements,
which is smaller than that of velocity shifts.
This implies that the Doppler shift is the major kinetic motion of these ejecta. 

The last two panels in Figure~\ref{fig:maps} show the difference between IME and Fe on Doppler shift (e) and Doppler broadening (f)
to emphasize the differences of these two elements.
In the panel (e), red and blue mean more blueshift and more redshift in Fe, respectively, and in the panel (f) they mean more broadened and less broadened in Fe, respectively.
The panel (e) is very noisy with the error range of a few 100s km~s$^{-1}$, which is due to the low statistics of the Fe-K map.
The absolute values of the shift looks larger in Fe,
but the large error regions of Fe fitting prevented us from concluding it.
It will be discussed in section ~\ref{sec:radial}.
The panel (f) shows that 
the northern part has larger velocity dispersion in IME than Fe although it is noisy with an error range of a few 100s km~s$^{-1}$,
which may indicate that the IME expansion in this region is faster than Fe,
or there may be more velocity components contributing to the IME velocity dispersion.

\section{Discussion}
\label{sec:discuss}

We have measured
the position dependence of the Doppler shift and the broadening of Fe,
and compared them with those of the IME as shown in Figure~\ref{fig:maps}.
Here we discuss the expansion structure of Cas~A with our measurements.


\subsection{Radial profiles of expansion}
\label{sec:radial}

\begin{figure*}
 \begin{center}
  \includegraphics[width=7cm]{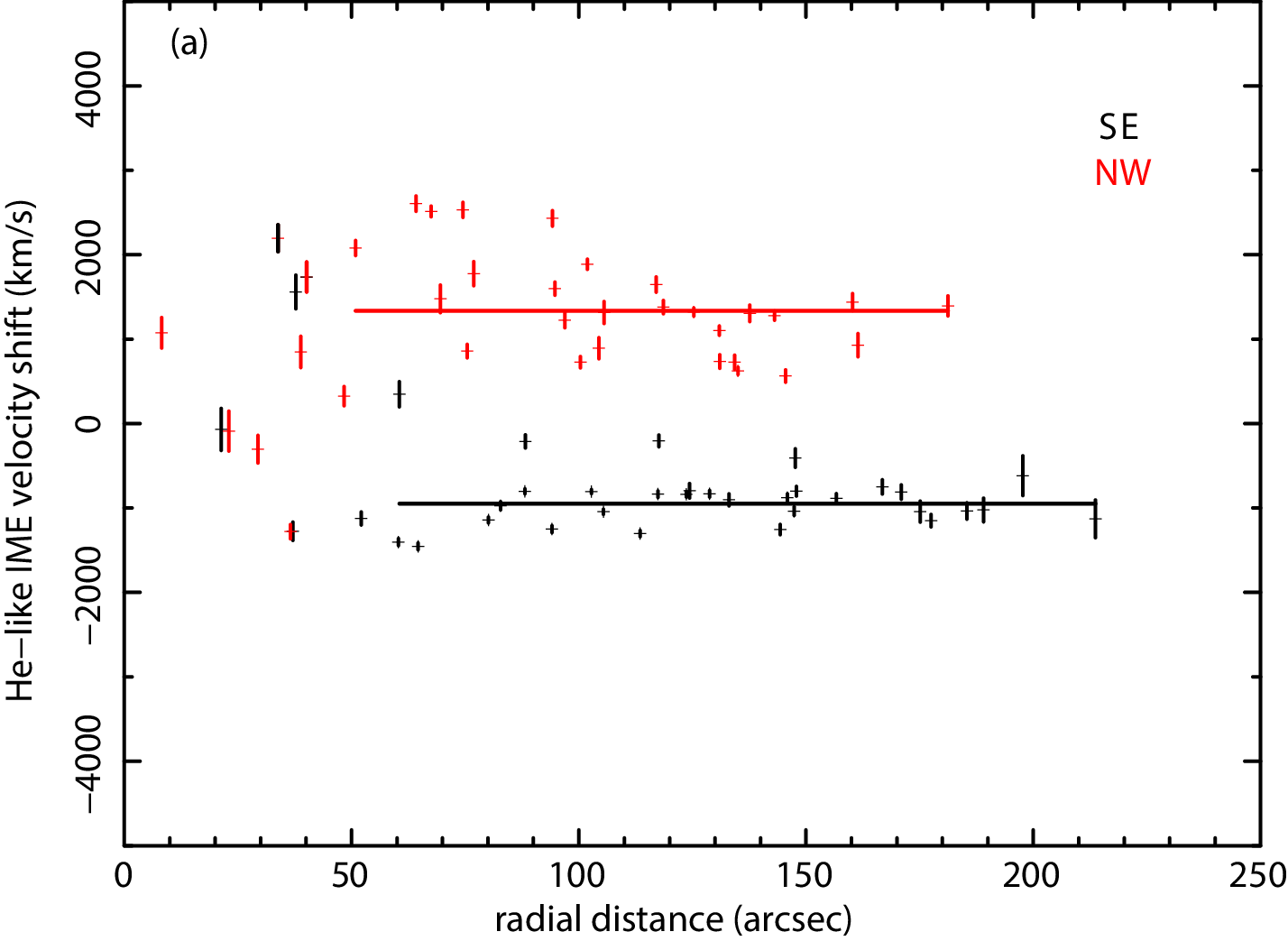} 
  \includegraphics[width=7cm]{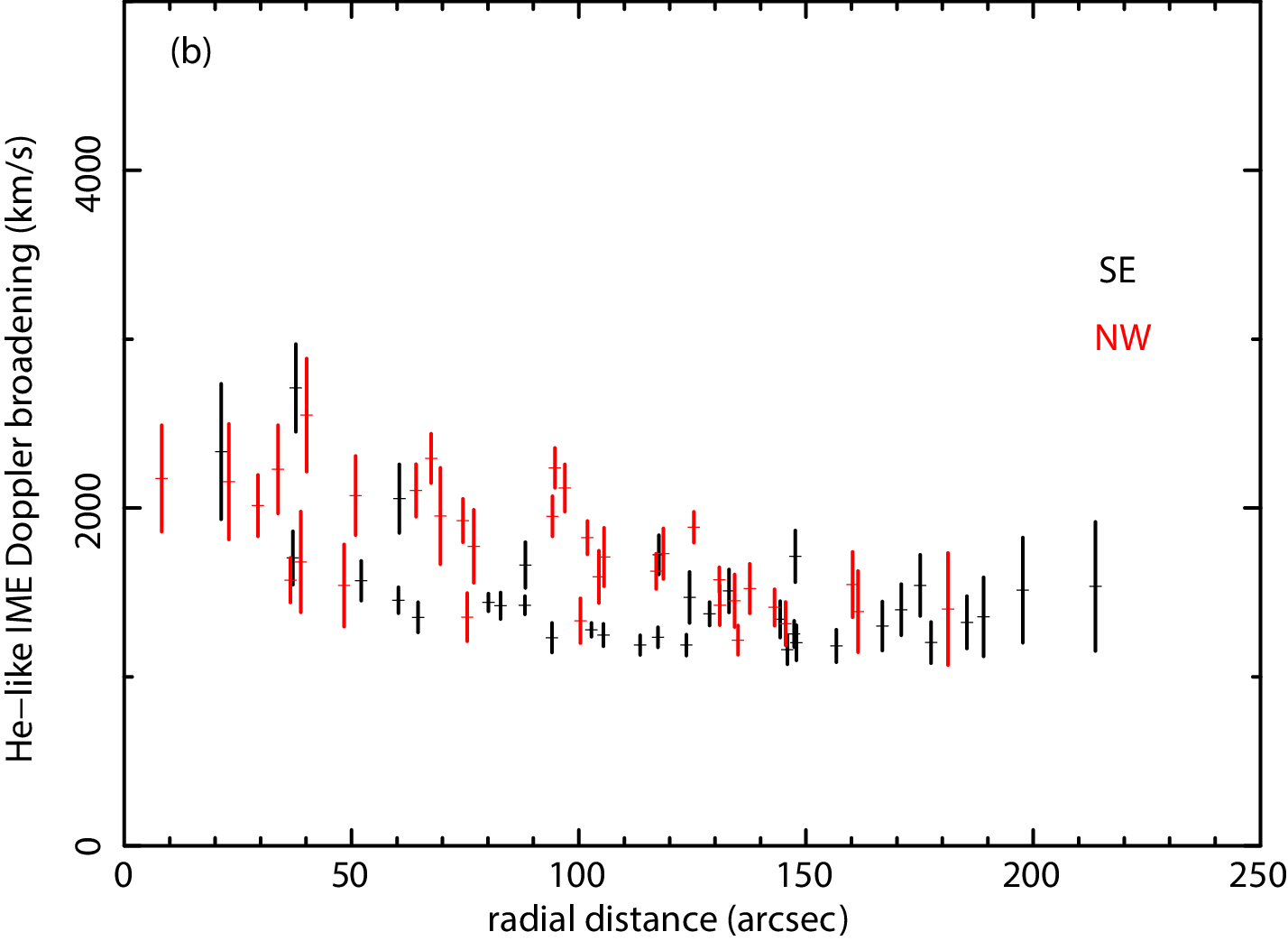} 
  \includegraphics[width=7cm]{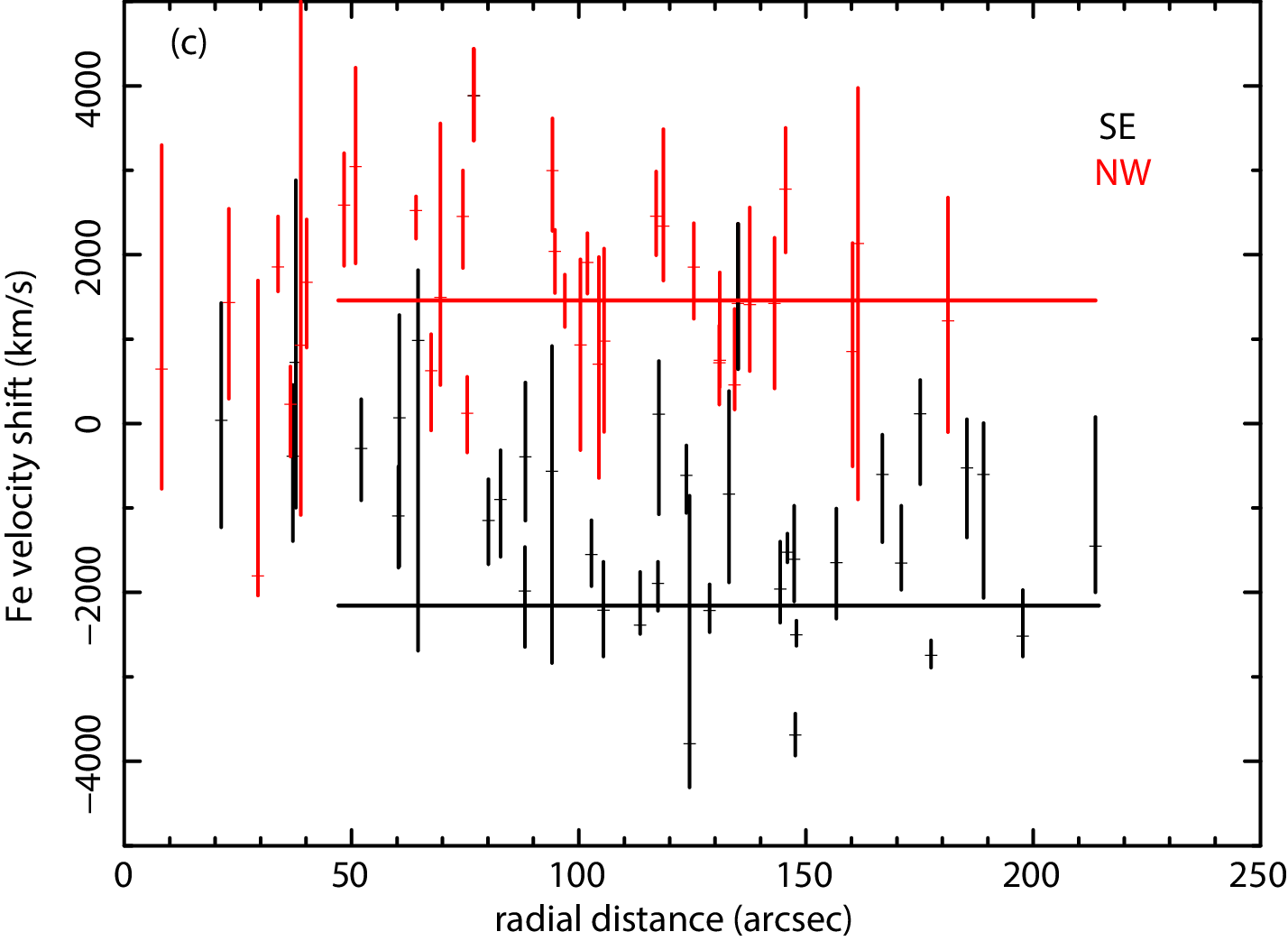} 
  \includegraphics[width=7cm]{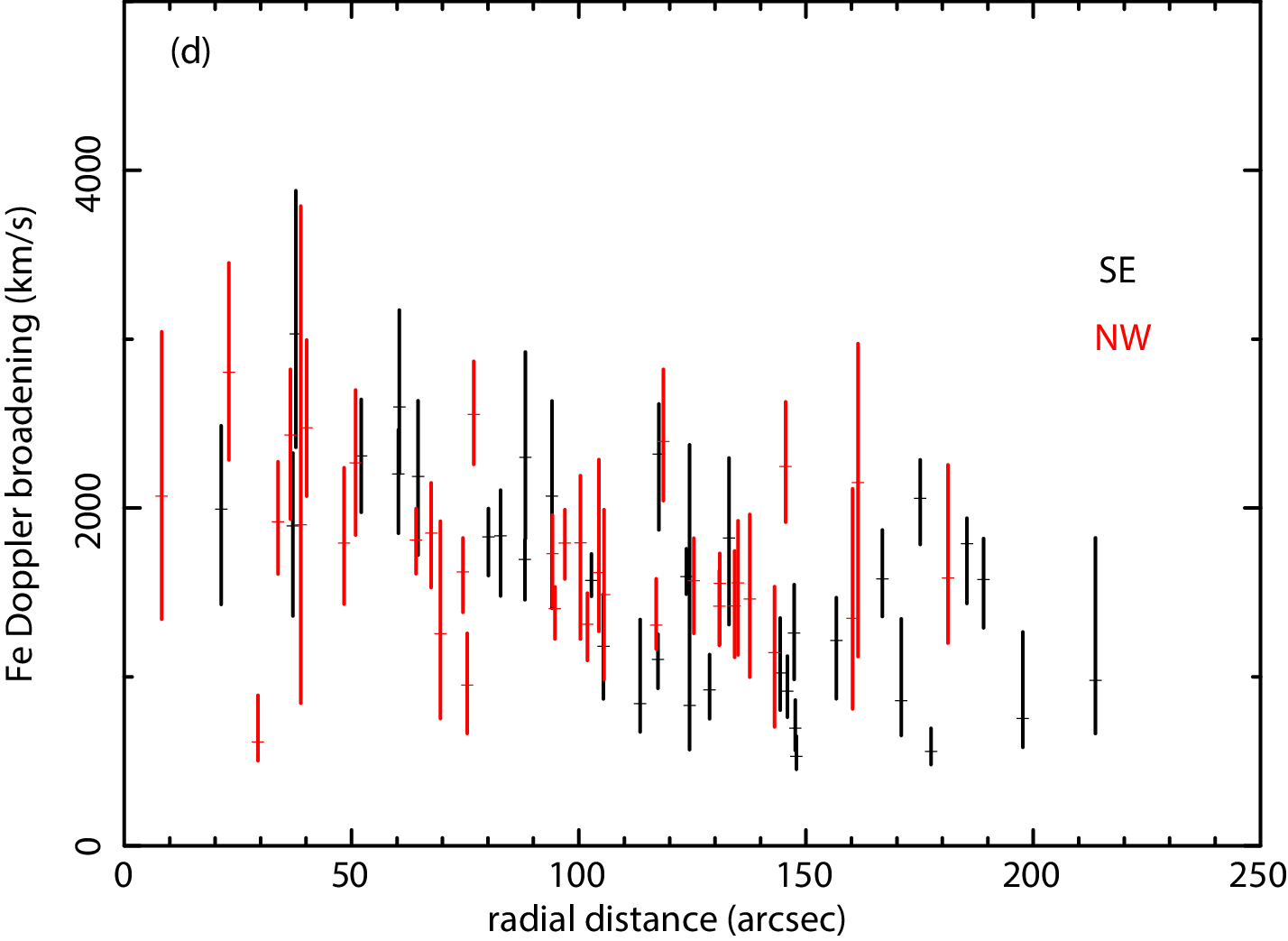} 
\end{center}
\caption{Radial profiles of (a) Doppler shift of He-like IME,
(b) Doppler broadening of He-like IME,
(c) Doppler shift of Fe-K,
and (d) Doppler broadening of Fe-K. 
The data for IME is taken from \citet{vink2024}.
Black and red lines in the panel (a) and (c) show the average values of Doppler velocity.
{Alt text: Four line graphs showing the radial profile of Doppler parameters.}
}\label{fig:radial}
\end{figure*}

In this subsection, we discuss the overall expansion structure of plasma
emitting He-like IME and Fe-K lines from the radial profiles of Doppler parameters.
Here, we treat the position (RA, Dec.) = (\timeform{23h23m27.77s}, \timeform{58D48'49.4''}) as the expansion center from \citet{thorstensen2001} (see Figure~\ref{fig:maps} (a)).
Figure~\ref{fig:radial} shows the radial profiles of the Doppler parameters.

Figure~\ref{fig:radial} (a) and (c) show the Doppler shift of He-like IME and Fe and radial distance, respectively.
It is clearly seen that the NW region shows redshift and the SE region shows blueshift, which represents asymmetric expansion.
Such a structure can be reallized
if the ejecta shows bipolar-like expansion, or the ejecta forms incomplete shell.
In the case of bipolar expansion,
the absolute value of the redshift should be larger in the regions distant from the center of the remnant,
but we do not see such a tendency.
We thus concluded that the ejecta forms an incomplete shell structure in both the IME and Fe.
One can also find that the absolute value of Doppler shift of IME is larger in the NW region, which is not clearly seen in Fe.
In order to check this,
we derived the average values of Doppler shift in the regions distant more than 50~arcsec from the expansion center.
The average absolute values of the Doppler shift velocity of IME are $\sim$950~km~s$^{-1}$ in SE and $\sim$1300~km~s$^{-1}$ in NW
with the error range of a few 10s~km~s$^{-1}$,
whereas those for Fe are  $\sim$2100~km~s$^{-1}$ in SE and $\sim$1400~km~s$^{-1}$ in NW with the error range of a few 100s~km~s$^{-1}$.
We cannot adress the difference in the asymmetry between Fe and IME
due to the large uncertainties.

Figure~\ref{fig:radial} (b) and (d) show the Doppler broadening versus radial distance of IME and Fe.
The broadening is larger at the small radial distance (center) region
in both elements,
implying the expansion of the ejecta.
These results also support our conclusion that the Doppler shift trend in panel (a) and (c) is not due to bipolar expansion but incomplete shell.
The broadening of IME is larger in the NW region than in the SE region \citep{vink2024}, which implies an asymmetric expansion.
This is also suggested by the absolute values of the Doppler shifts of the IME. 
The absolute values of the Doppler shifts of Fe also indicate asymmetric expansion,
although the large statistical errors do not allow us to conclude which element shows more asymmetry.
The expansion velocities are
2000--3000~km~s$^{-1}$,
which are well below the average forward shock velocity (5800~km~s$^{-1}$; \cite{vink2022}),
and align quite well with 
those expected at the age of
Cas A for the remnant of a neutrino-driven supernova explosion (see the middle panels in Fig. 6 of \cite{orlando2021}).

\citet{willingale2002} claimed that there are some regions with higher positive (redward) velocity shifts of Fe compared with those of the IME,
with values larger than $\sim$2000~km~s$^{-1}$.
\citet{hughes2000} revealed that
the Fe ejecta in the southeastern region are in more outer region than IME
(see also \cite{hwang2012,sato2021}).
In our analysis, the absolute values of velocity shifts are larger than 2000~km~s$^{-1}$ in several pixels,
although the relatively large errors do not allow us to definitively conclude that the Fe has a higher velocity than the IME in these regions.
Nevertheless, this analysis suggests that Fe indeed exhibits higher velocities than IME in the SE region and possibly in part of the NW region, although the uncertainties prevent us from drawing a definitive conclusion. This finding provides further evidence of the large-scale asymmetries resulting from the SN explosion.

Note that the half-power diameter of the point spread function (PSF) of the X-ray Mirror Assembly (XMA) onboard XRISM is larger than the Resolve pixel size
\citep{boissay-malaquin2022,tamura2022,hayashi2022},
and we have to consider contamination of photons
from nearby pixels, when we analyze single-pixel spectra
\citep{plucinsky2024}.
This effect may dilute the spatial structure of the Doppler parameters on scales smaller than or comparable to the PSF size.
The impact will be greater if the radiation of the pixel of interest is smaller than that of the surrounding pixels.
On the other hand, 
it should not have a significant effect on structures that are larger than the PSF size,
such as radial profiles we derived.

\subsection{Regions with narrow Fe lines}

\begin{figure}
 \begin{center}
\includegraphics[width=7cm]{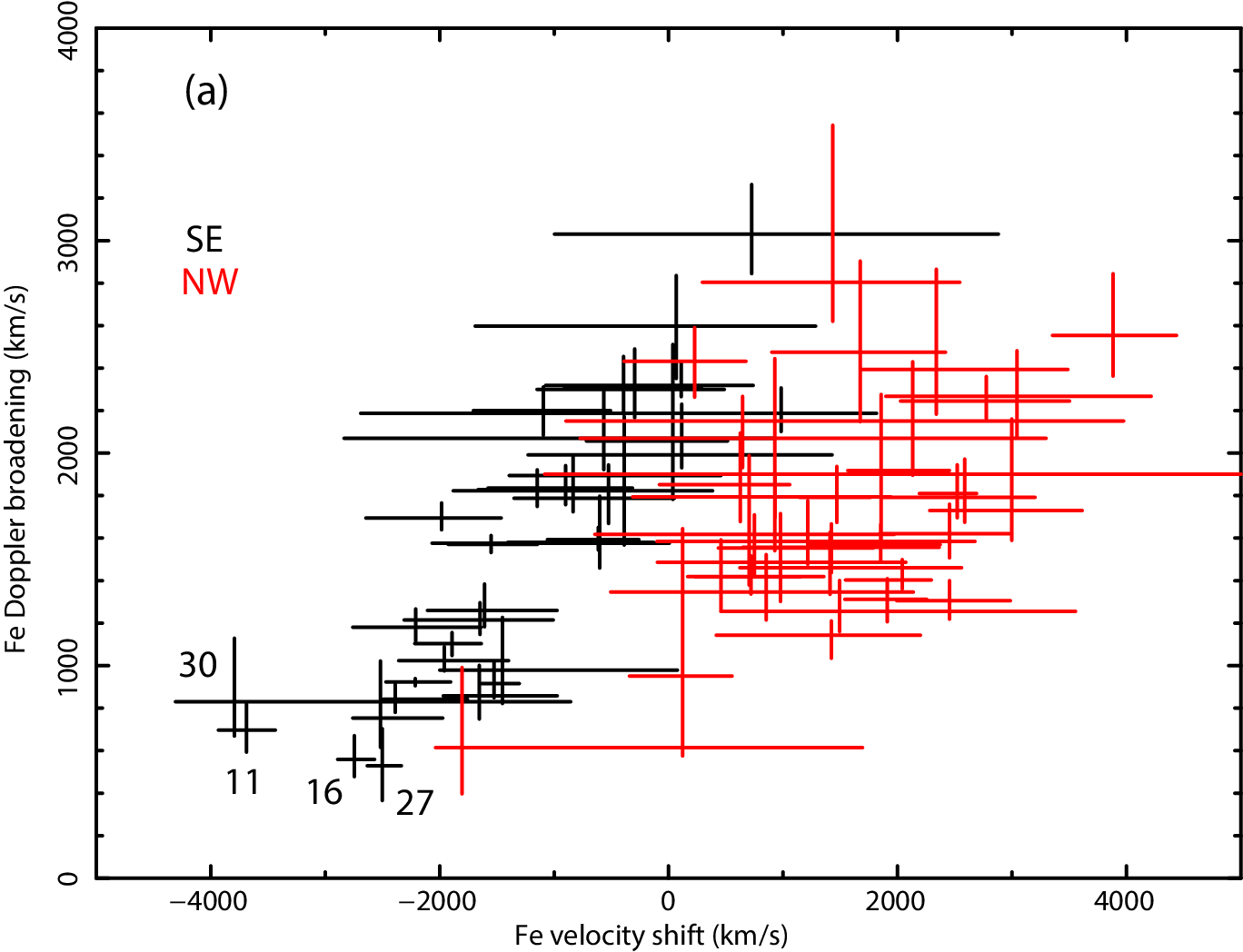}
  \includegraphics[width=7cm]{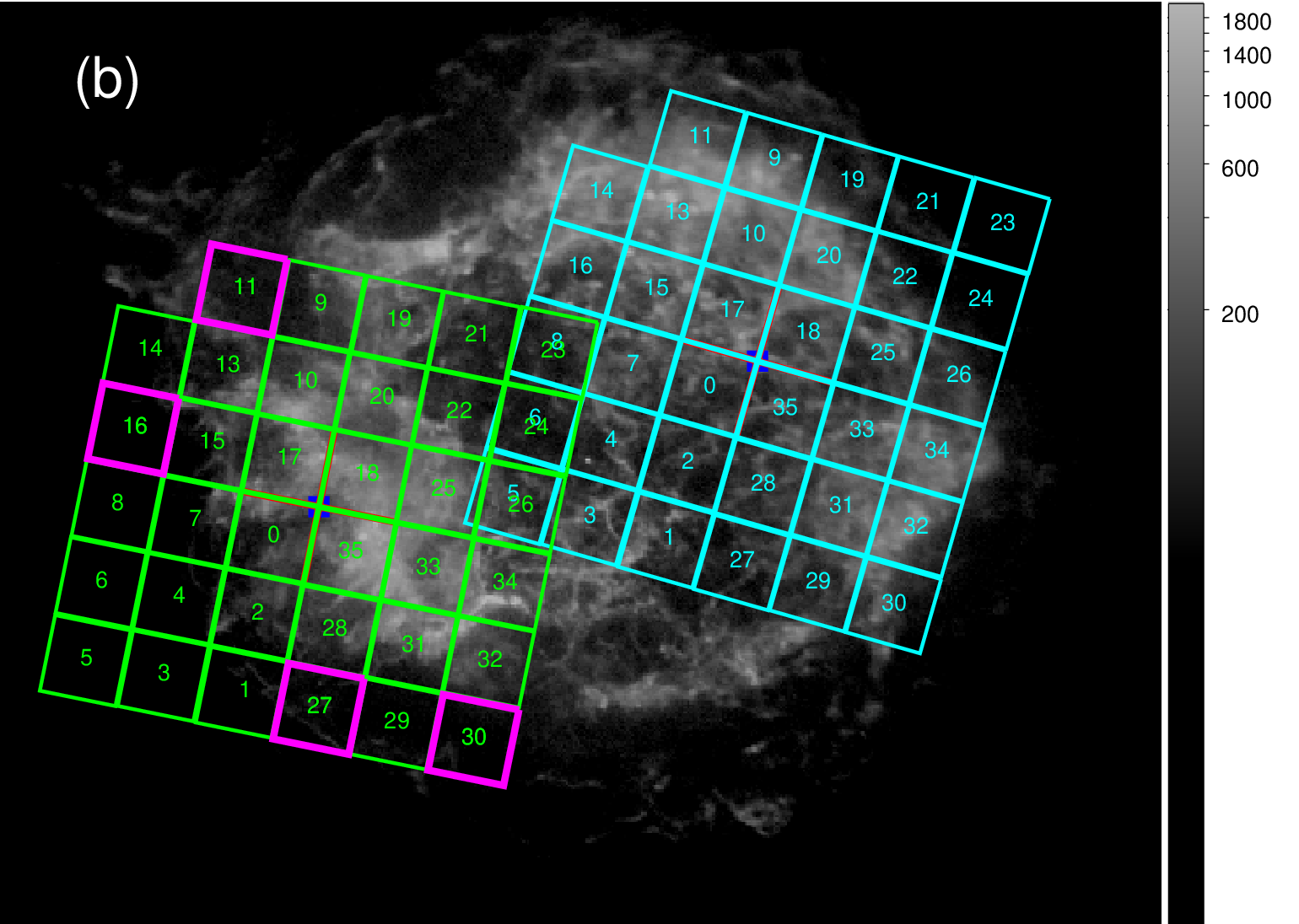}
\end{center}
\caption{Panel (a): Doppler broadening of Fe vs. the redshift.
Black and red data represent the data from SE and NW regions, respectively.
The numbers in the figure are pixel numbers.
Panel (b):
All-band Chandra image in the logarithmic scale (black and white) with Resolve field of view.
Magenta boxes represent the pixels with blueshifted and narrow Fe lines. 
{Alt text: A Line graph showing Doppler broadening of Fe vs. the redshift, and a map showing the blueshifted and narrow Fe line region on Cas A.} 
}\label{fig:v-w}
\end{figure}

\begin{figure*}
 \begin{center}
  \includegraphics[width=7cm]{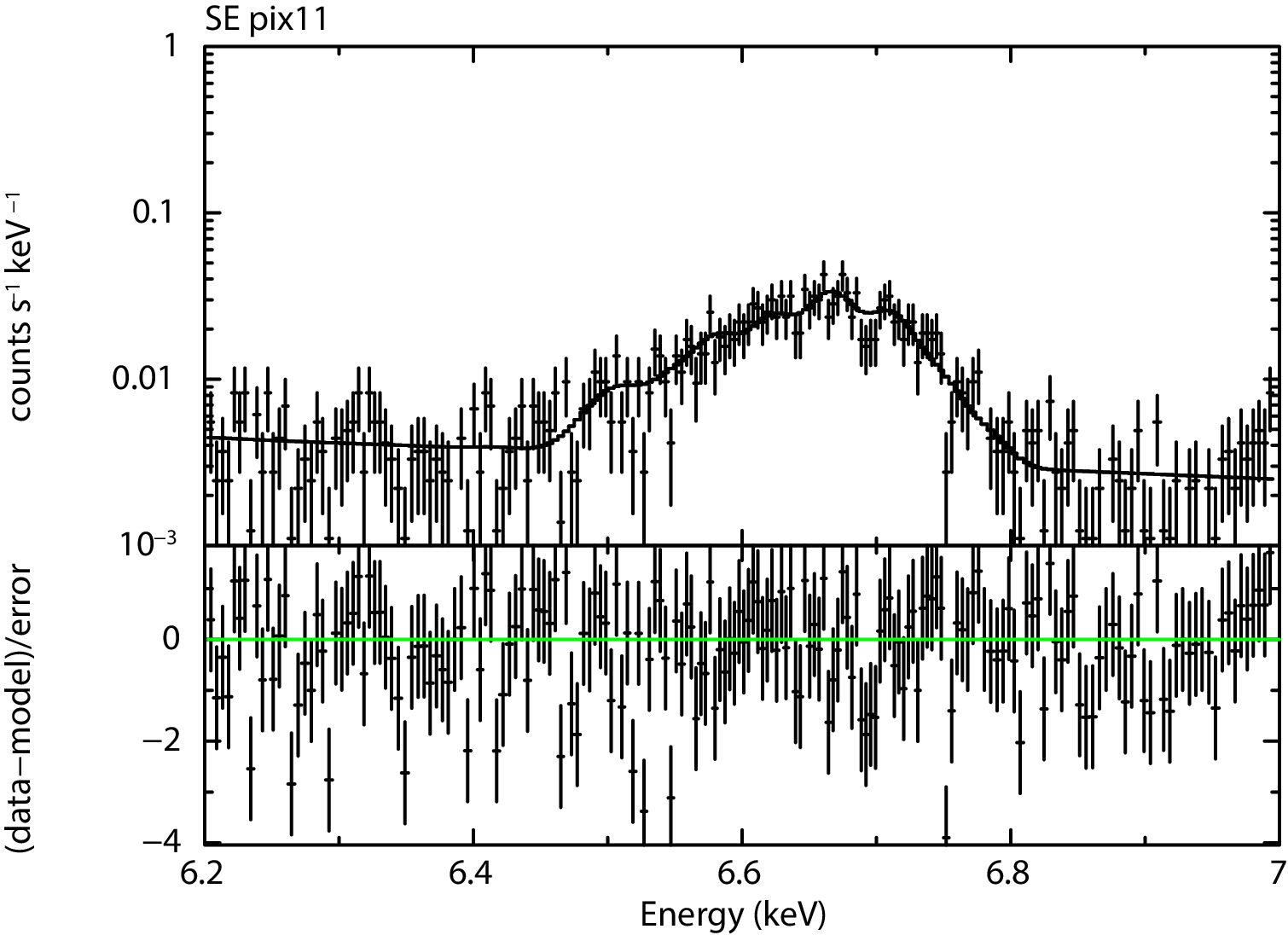}
  \includegraphics[width=7cm]{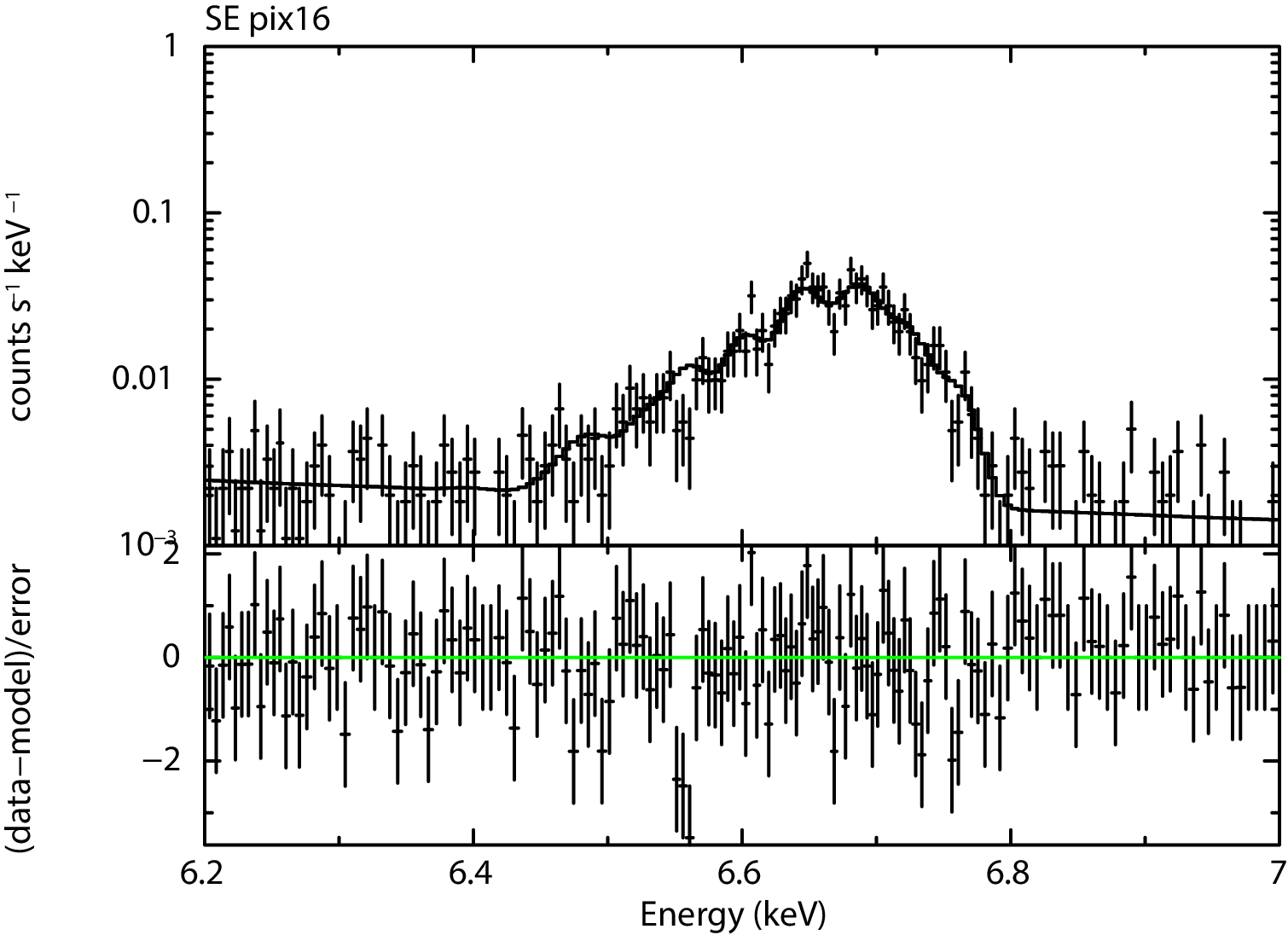}
  \includegraphics[width=7cm]{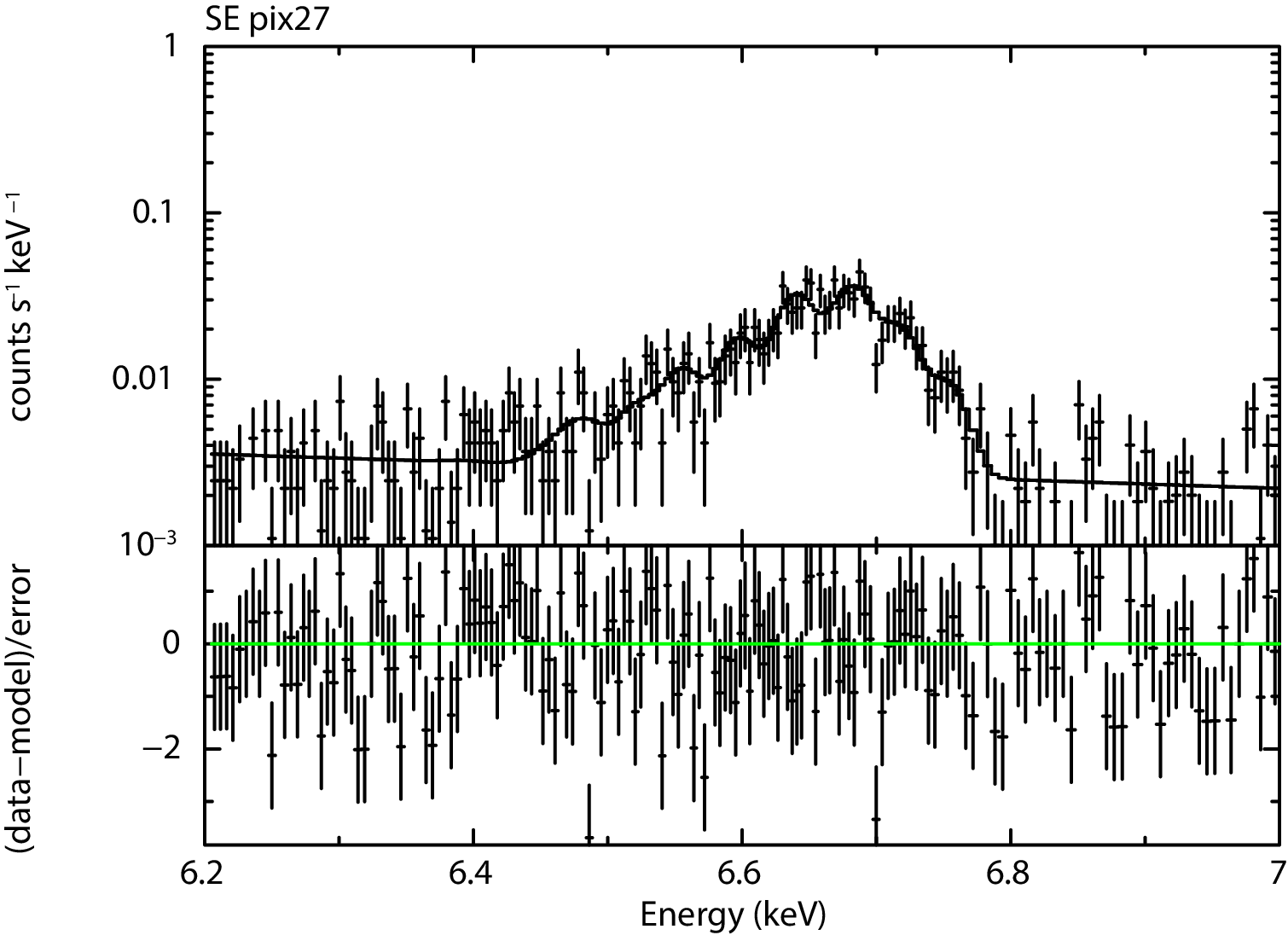}
  \includegraphics[width=7cm]{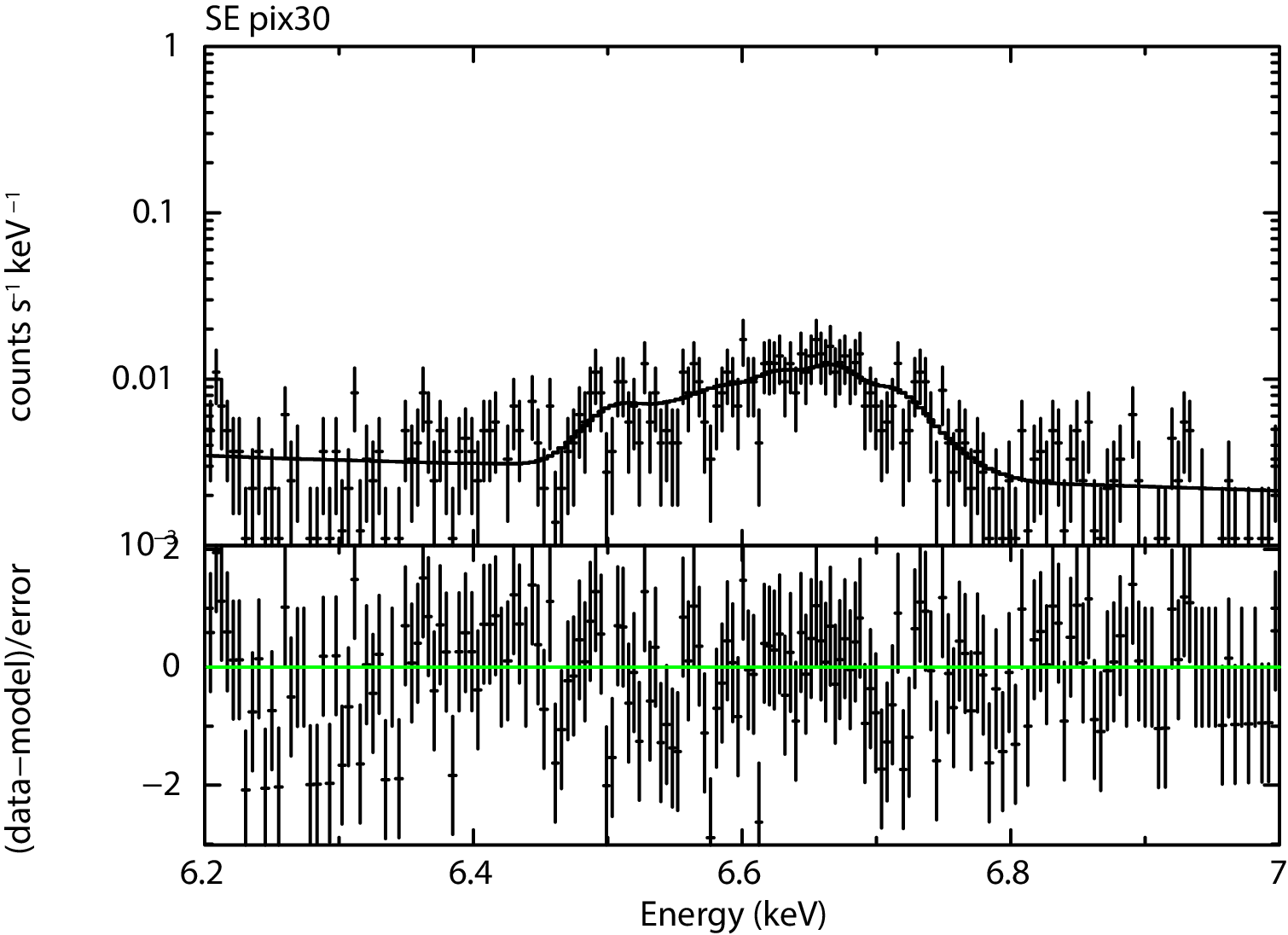}
 \end{center}
\caption{Fe K-shell line feature with the pixel 11, 16, 27, and 30 in the SE observation,
which shows small line broadening.
The solid and dotted lines in each subfigure represent the sum and individual components in the fitting.
The bottom panel in each subfigure shows the residuals from the best-fit model.
{Alt text: Four line graphs showing the Fe line feature of pixel 11, 16, 27, and 30.}
}\label{fig:FeLines}
\end{figure*}

In this subsection, we discuss the smaller scale asymmetry of the Fe Doppler parameters.
Figure~\ref{fig:v-w} (a) shows redshift vs. Doppler broadening of the Fe line.
In the case of uniform expansion of a uniform sphere,
the redshift should be 0 in the entire remnant if the emission is integrated along the line of sight, and its deviation implies divergence from isotropy.
One can see that
the spectra from several pixels show very blueshifted and narrow Fe lines.
This implies that the Fe ejecta in these regions are really moving towards the oberver with small dispersion.
Among them, we selected four examples with Doppler broadening below 850~km~s$^{-1}$ and Doppler shift below $-2300$~km~s$^{-1}$,
pixel 11, 16, 27 and 30, as highlighted in Figure~\ref{fig:v-w} (b).
Figure~\ref{fig:FeLines} presents the spectra in the 6.2--7~keV band from these pixels.
There is suggestive substructure on the top of the Fe line bump.
This may be due to the reduced line blending in this region.
Panel (b) of Figure~\ref{fig:v-w} shows the position of these pixels. All of pixels with narrow Fe lines are on the edge of the eastern half of the remnant.

Such narrow and blueshifted lines on the edge of the eastern half of the remnant shows that the Fe ejecta in this region moving towards us.
On the other hand, we have no narrow line region in the NW region, suggesting that there is no ejecta moving away from us.
These results 
imply that the ejecta form an incomplete shell.
The blueshifted Fe ejecta in the east of the remnant was previously suggested by \citet{willingale2002},
and we have confirmed the scenario that the Fe ejecta in this area is blueshifted with low dispersion.

Pixel~11, which is on the root of the northeastern jet,
is especially interesting in the context of the jet motion.
This region has narrow Fe lines, whereas He-like IME shows broad feature (see Figure~\ref{fig:maps} (b)).
In this region, O-rich ejecta detected by HST and S-rich ejecta also show high velocity proper motion \citep{fesen2006b,fesen2006a}.
Their velocity is $\sim$6,000--14,000~km~s$^{-1}$, which are much faster than
that of the Fe ejecta we measured.
Thus the Fe ejecta in this region have a different motion
from that of S-rich ejecta and debris forming the jet,
or the jet does not contain Fe-rich ejecta.
These results suggest that Fe-rich ejecta do not penetrate the progenitor’s outer layers of lighter elements in this jet direction.

The broadening on the edge of the remnant contains thermal broadening.
Assuming that the broadening discussed here, 600--800~km~s$^{-1}$, is purely due to thermal broadening, 
the Fe temperature is estimated to be 250--510~keV.
This is a kind of the upper limit of Fe temperature in this region,
since there should be some contamination of the kinetic Doppler effect.
With further assumption that the Fe ejecta are ideal gas,
the reverse shock velocity which heated the ejecta are derived to be 1690--2550~km~s$^{-1}$
from Rankine-Hugoniot relation.
The reverse shock velocity can be very low or very high, depending on the relative velocity of the reverse shock with reference to the unshocked ejecta \citep{vink2022}.
Other possibilities of the rather low velocities are that the particle acceleration is efficient which makes the deviation from the ideal gas assumption,
or partial equilibration discussed in the N132D case \citep{xrism2024}.
Further analysis is needed in the future studies.

Note that there should be the spatial-spectral mixing due to the PSF, which is mentioned in \citet{plucinsky2024}.
Spatial-spectral mixing can not artifically narrow line emission in a complex three-dimensional structure like Cas A, it can only broaden the lines.  Therefore, if narrow lines are observed the adjacent regions must have similarly narrow lines and/or the effects of spatial-spectral mixing must be small.
Note also that the line energy shift and the broadening width couple with the ionization time scale.
See the appendix for the details.

\subsection{On the explosion mechanism of Cas A}


The expansion structure of Fe and IMEs contains important information on the explosion mechanism of Cas A.
Overall, 
the northewestern part shows a redshift and the southeastern part a blueshift for both IME and Fe, indicating an incomplete shell structure.
IME ejecta expand faster in the northwestern region,
whereas the Fe ejecta does not show such a trend significantly.
For the smaller scale,
the jet region in the northeast shows broad IME lines and narrow Fe lines,
indicating the jet does not contain Fe.
In this subsection, we discuss the possible scenario to explain our observations.

Incomplete shocked shell structures rich in IME and Fe-group elements can naturally arise from a neutrino-driven supernova (SN) explosion.
\citet{orlando2021} presented three-dimensional magnetohydrodynamic simulations describing the evolution of a neutrino-driven SN, from the explosion to the fully developed remnant at the age of Cas~A. The simulations begin with initial conditions provided by a neutrino-driven SN model characterized by three extended Ni-rich plumes, oriented in directions consistent with the Fe-rich regions observed in Cas A \citep{wongwathanarat2017}.
These large-scale asymmetries in the Ni distribution form self-consistently within the first seconds after core collapse, driven by convective overturn due to neutrino heating and the standing accretion shock instability \citep{wongwathanarat2015,wongwathanarat2017}
The nickel in these plumes decays into Co and subsequently into Fe during the first year following the shock breakout.
Such an asymmetry in neutrino-driven SNe are also shown in calculations by \citet{burrows2024} and \citet{vartanyan2025}, for example.

According to the model by \citet{orlando2021},
at approximately 30 years post-explosion, the Fe-rich plumes begin interacting with the reverse shock, initiating the formation of the Fe-rich regions observed in Cas A. For this reason, the Fe-rich regions can be considered a direct signature of the explosion dynamics. By the age of $\sim$200~years, the remnant encounters an asymmetric, dense circumstellar shell, further amplifying asymmetries in the remnant's structure \citep{orlando2022}. These models suggest that the Fe-rich regions observed in Cas A today correspond to the shocked heads of the original Ni-rich plumes, which were formed as a direct result of the neutrino-driven SN explosion. 
According to the model, these regions form incomplete shocked shell structures rich in IMEs and Fe-group elements:
one redshifted in the NW hemisphere and the other blueshifted in the SE hemisphere, as revealed by XRISM/Resolve in Cas A.
These incomplete shells, therefore, may correspond to those driving the observed asymmetry in the velocities of IMEs and Fe detected by XRISM across the SE and NW regions of the remnant,
further supporting a neutrino-driven mechanism for the SN explosion.

\section{Conclusion}
\label{sec:conclusion}
It is investigated the ejecta expansion structure of the young core-collapse SNR Cas~A,
with the Doppler parameter mapping of Fe-K line with the excellent energy resolution of XRISM/Resolve.
It is found that the Fe ejecta expand asymmetrically,
with blueshifted in southeast and redshifted in northwest,
similar to the IMEs.
The line broadening is larger in the center with the value of $\sim $2000--3000~km~s$^{-1}$, and smaller on the edge of the remnant.
We found highly blueshifted, and narrow Fe-K lines on the SE region
although the significance is not great,
whereas no such narrow line was found in the NW region.
These features in larger and smaller scales may indicate an incomplete shell structure.
These findings are consistent with the asymmetries predicted by models of neutrino-driven supernova explosions \citep{wongwathanarat2017,orlando2021}.
%
%

\begin{ack}
We would like to thank the anonymous referee for his/her comments.
We thank all the XRISM members.
AB thanks Keiichi Maeda, Shiu Hang Lee, and Susumu Inoue for their discussion.
This work was financially supported the JSPS Core-to-Core Program (grant number: JPJSCCA20220002),
Japan Society for the Promotion of Science Grants-in-Aid for Scientific Research (KAKENHI) Grant Numbers, JP23H01211, JP23K25907 (AB), and JP20K04009 (YT).
The work of JV and MA on this paper is  (partially) funded by NWO under grant number 184.034.002.
SO acknowledges financial contribution from the PRIN 2022 (20224MNC5A) - "Life, death, and after-death of massive stars," funded by the European Union – Next Generation EU.
\end{ack}





\section*{Appendix: Fitting consistency check with ionization time scale map}
\label{sec:nt}

\begin{figure}[hbtp]
\begin{center}
\includegraphics[width=8cm]{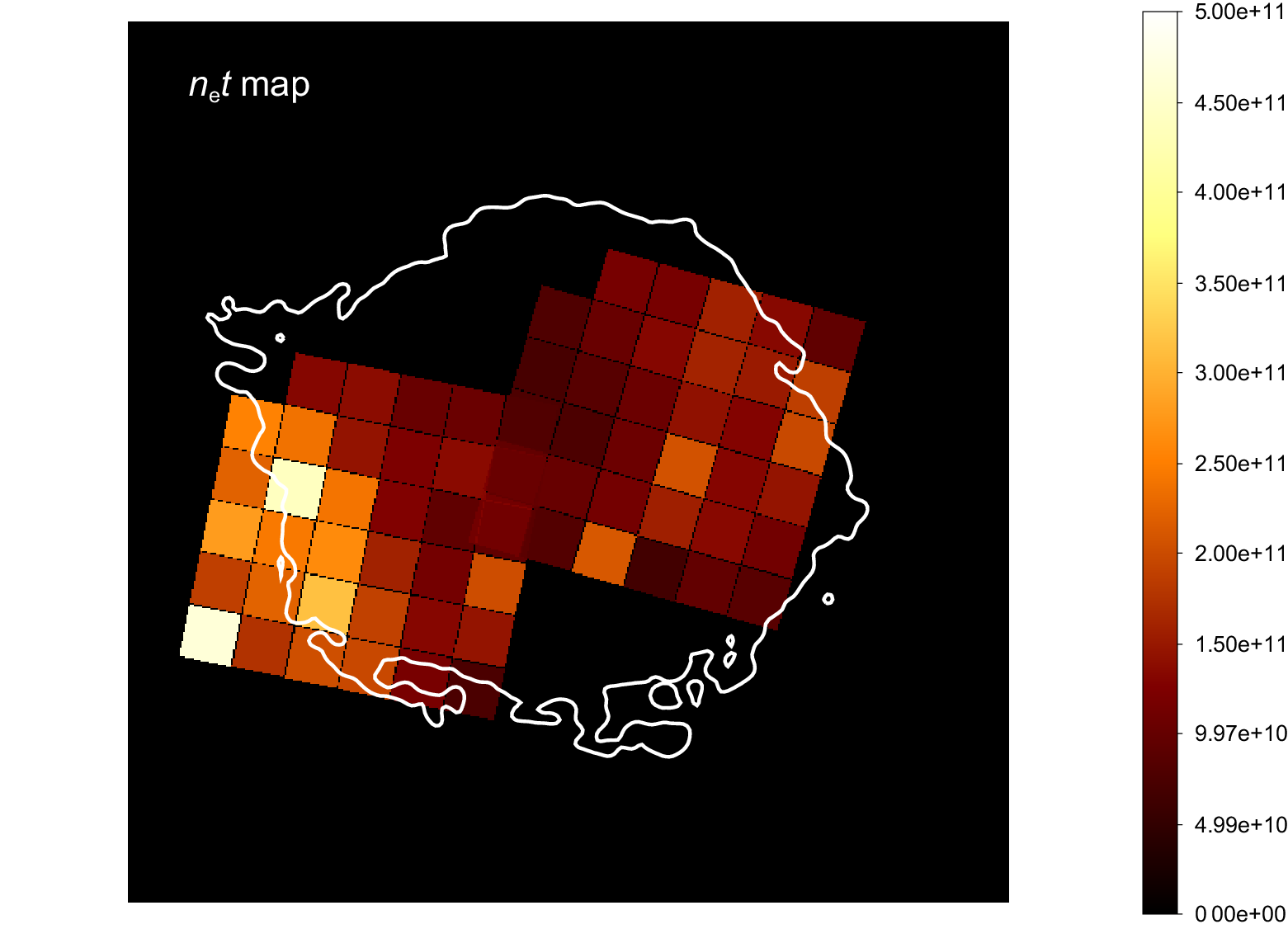}
\end{center}
\caption{Map of ionization time scale, $n_et$, }of the Fe component. The color scale is in the linear scale in the unit of cm$^{-3}$s.
The white contour shows the outer boundary of the remnant from the Chandra whole-band image.
{Alt text: A map of net of Fe-K component.}
\label{fig:nt}
\end{figure}

For the consistency check and examining the dependency of our main results from our fittings, we made the ionization time scale ($n_et$) map as shown in Figure~\ref{fig:nt}.
Note that 
the $n_et$ discussed here is the one derived with the spectral model {\tt pshock}, and therefore the maximum value in the plasma.
We compare it with 
the $n_et$ map made by the wide-band fitting with Chandra \citep{hwang2012}.
Our map approximately follows the trend of Figure~1 in \citet{hwang2012};
the $n_et$ is smaller in the center and larger in the outer regions, which is consistent with the expectation that the Fe ejecta are heated by the reverse shock.
The southeastern regions have the largest $n_et$ on the order of a few  $\times 10^{11}$~cm$^{-3}$s,
which is also consistent with the result that the IME have higher expansion velocity in NW \citep{vink2024},
implying that the density in the SE region is higher.
This is also well consistent with the previous results \citep{hwang2012}. 

\begin{figure}[hbtp]
\begin{center}
\includegraphics[width=7cm]{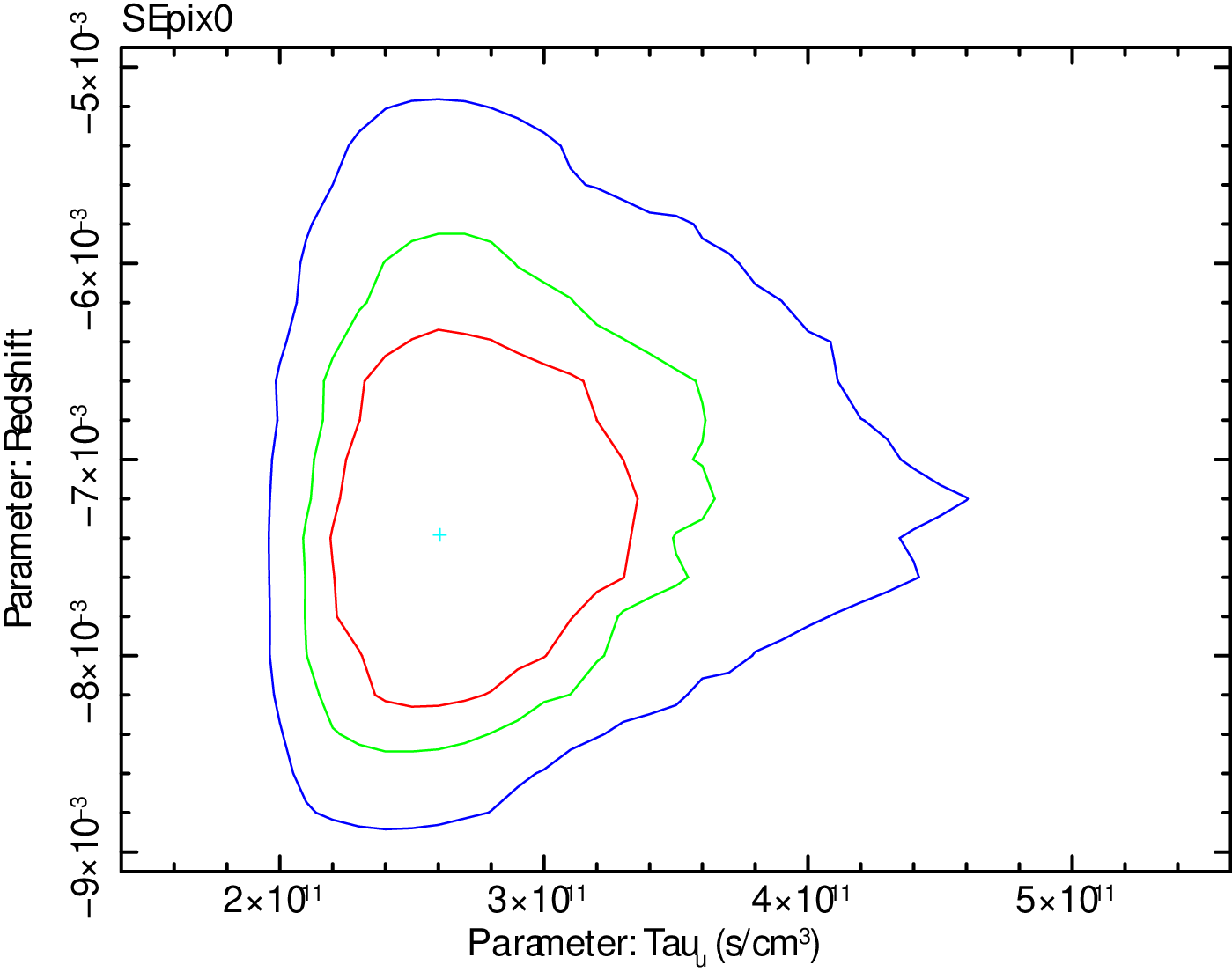}
\includegraphics[width=7cm]{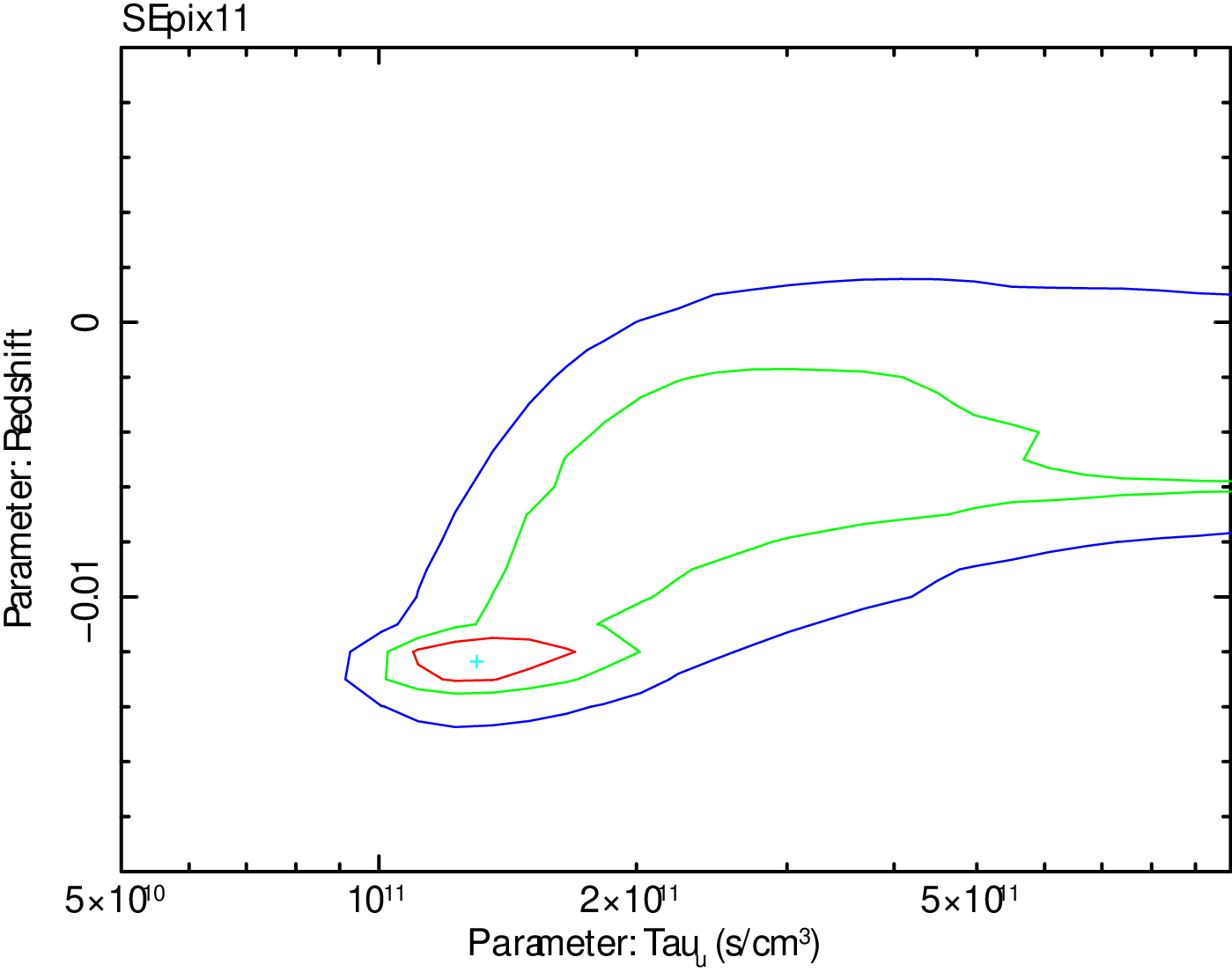}
\includegraphics[width=7cm]{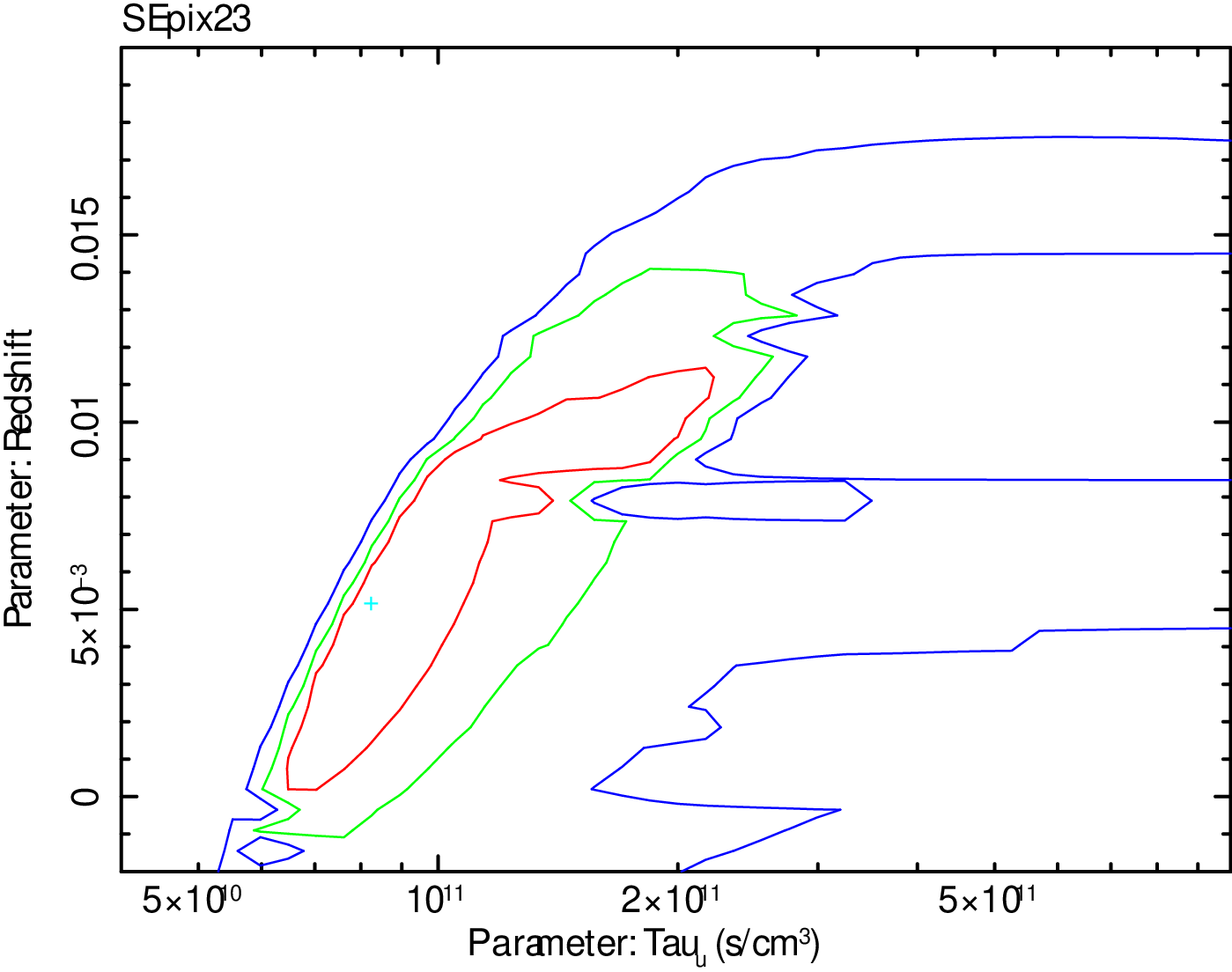}
\end{center}
\caption{Confidence contour maps between the $n_et$ in the unit of cm$^{-3}$s and the velocity shift (in dimensionless redshift) for the pixel 0 (top), pixel 11 (middle), and pixel 23 (bottom) of the SE observation.
The red, green, and blue contours represent 68\%, 90\%, and 99\%, respectively.
{Alt text: Three plots of confidence contours between nt and velocity shift for pixel 0, 11, and 23 of the SE observation.}
}\label{fig:contours}
\end{figure}

On the other hand, our best-fit parameters have large error regions and sometimes dependencies on other parameters.
Figure~\ref{fig:contours} shows the confidence contour plots between $n_et$ and the velocity shift of pixel 0, 11, 23 of the SE observation
(see \citet{plucinsky2024} for reference to the number of pixels, or Figure~\ref{fig:v-w}).
The top panel, pixel 0, which is the typical region with high $n_et$,
shows that the $n_et$ and the velocity shift do not correlate with each other.
In such regions, 
the maximum ionization state of Fe reaches He-like and the charge-state distribution of the Fe ejecta modeled by the plane-parallel shock has a broad range of ionization states. Because of this, the broadened line profile becomes highly asymmetric.
As a result, the $n_et$ and the velocity shift are almost uncorrelated.
The middle and botom panels show the cases with lower $n_et$.
In this case, the two parameters show some correlation.
In our results, 68--90\% error regions are rather well determined, but we have very large 99\% error range.
Although we also tried fitting including H-like Fe band,
no crucial difference was found.
Fitting in a larger energy band will be the subject of a future paper.

%
%


\end{document}